\documentclass{emulateapj}
\usepackage{natbib}
\usepackage{graphicx}
\usepackage{epstopdf}
\usepackage{amsmath}
\usepackage{subfigure}

\shorttitle{Finding Open Cluster Properties}

\begin{document}

\title{Open Clusters as Probes of the Galactic Magnetic Field: I. Cluster Properties}

\author{Sadia Hoq\altaffilmark{1} and D. P. Clemens\altaffilmark{1}}
\altaffiltext{1}{Institute for Astrophysical Research, 725 Commonwealth Ave, Boston University, Boston, MA 02215; shoq@bu.edu, clemens@bu.edu}

\begin{abstract}
Stars in open clusters are powerful probes of the intervening Galactic magnetic field, via background starlight polarimetry, because they provide constraints on the magnetic field distances.  We use 2MASS photometric data for a sample of 31 clusters in the outer Galaxy, for which near-IR polarimetric data were obtained, to determine the cluster distances, ages, and reddenings via fitting theoretical isochrones to cluster color-magnitude diagrams.  The fitting approach uses an objective $\chi^2$ minimization technique to derive the cluster properties and their uncertainties.  We found the ages, distances, and reddenings for 24 of the clusters, and the distances and reddenings for six additional clusters that were either sparse or faint in the near-IR.  The derived ranges of log(age), distance, and E$(B-V)$ were 7.25-9.63, $\sim$670--6160 pc, and 0.02-1.46 mag, respectively.  The distance uncertainties ranged from $\sim$8 to 20\%.  The derived parameters were compared to previous studies, and most cluster parameters agree within our uncertainties.  To test the accuracy of the fitting technique, synthetic clusters with 50, 100, or 200 cluster members and a wide range of ages were fit.  These tests recovered the input parameters within their uncertainties for more than 90\% of the individual synthetic cluster parameters.  These results indicate that the fitting technique likely provides reliable estimates of cluster properties.  The distances derived will be used in an upcoming study of the Galactic magnetic field in the outer Galaxy.
\end{abstract}

\section{Introduction}
Open star clusters in the Galactic plane are useful tools to probe Galactic properties.  Because cluster members are generally coeval \citep{Trumpler1925, Friel1995}, they are roughly the same age and located at the same distance.  These properties make them excellent potential probes of the Galactic magnetic field via background starlight polarimetry \citep[e.g.,][]{Hall1949a, Hiltner1949, Serkowski1965}.  The measured linear polarization orientation of background starlight traces the direction of the plane-of-sky component of the magnetic field located between the observer and the stars.  The distance estimates provided by clusters can be used as upper limits to the proximity of the magnetic field along the line of sight.



The overarching goal of the study is to explore the large-scale Galactic magnetic field in the outer Galaxy, especially how its behavior changes with distance and in the presence or absence of spiral arms.  Using near-infrared (NIR) linear polarimetric observations of a sample of star clusters, we probed the magnetic field toward the outer Perseus arm and its inter-arm regions.  To find magnetic field properties in the interstellar medium (ISM) along these directions, however, we must establish cluster distances.  In this work, we compare background-subtracted cluster color-magnitude diagrams (CMDs), created using NIR 2MASS photometry \citep{Skrutskie2006}, to theoretical isochrones using a $\chi^2$ minimization approach to determine distances to the clusters, as well as their ages and reddenings.  The cluster distances derived through this approach will be the basis for an upcoming study of the Galactic magnetic field in the outer Galaxy (Hoq et al. 2016, in preparation).  

The cluster properties, including distances, of this sample have been determined in previous studies.  However, these studies used different datasets and different fitting techniques, and so have non-uniform systematics and uncertainties.  In addition, uncertainties of the cluster properties were often not reported.  Therefore, we developed an approach to fit isochrones to establish cluster properties and their uncertainties and applied it to a uniform dataset.

Several recent studies, using optical and NIR photometric data sets, have developed objective methods to fit isochrones to cluster CMDs with the goal of establishing cluster parameters, including, but not limited to, \citet{Naylor2006}, \citet{VonHippel2006}, \citet{Hernandez2008}, \citet{Monteiro2010}, \citet{Maciejewski2007}, \citet{Alves2012}, \citet{Dias2012}, \citet{Curtis2013}, \citet{Janes2013}, \citet{Lee2013}, and \citet{Perren2015}.  While all of these studies examined multiple clusters, their samples had little overlap with our sample of clusters (Section \ref{observations}).  Therefore, we developed an objective method to find the properties of the present sample of clusters.  

The method presented here shares similar features to many of the studies listed above, such as searching a parameter space consisting of age, distance, and reddening\footnote{We also briefly explored the effects of varying metallicity as a parameter.} by comparing stellar isochrones to cluster CMDs, but differs in other respects.  For example, the present study does not employ a Bayesian approach to search through the parameter space \citep[e.g.,][]{Monteiro2010, Alves2012, Janes2013, Perren2015}, but searches the complete parameter space.  In addition, this technique allows the derivation of the cluster properties without the need to identify which stars in the cluster field of view are cluster members and which belong to the field.  This was essential, as establishing cluster membership selection based solely on photometric data can be challenging.


The sample consists of 31 clusters in the outer Galaxy for which NIR polarizations were obtained.  Seven clusters failed the initial cluster CMD-isochrone fitting procedure because they were either too faint or too sparse.  Of these seven, the distances and reddenings were successfully derived for six clusters by fixing their ages.  For the 30 clusters that were fit, we found a median log(age) of $\sim$9.2, a median distance of 2900 pc, and a median E$(B-V)$ of $\sim$0.5 mag.

The structure of the paper is as follows.  Section 2 describes the cluster sample and the 2MASS NIR photometric data.  Section 3 describes the steps of the analysis, the results returned by the fitting procedure, and the tests of the fitting technique.  Section 4 discusses the distribution of the derived cluster properties and their comparisons to previous published results.  Our findings, summarized in Section 5, establish distances with 8--20\% uncertainties.

\section{Cluster Sample Selection and Observations}\label{observations}
To study the large-scale structure of the magnetic field, open clusters spanning a wide range of distance estimates and Galactic longitudes in the outer Galaxy were chosen for NIR polarimetric observations with the Mimir instrument \citep{Clemens2007}.  The sample of 31 chosen clusters spans Galactic longitudes $l\sim$119--232$^{\circ}$ and latitudes $b\sim -$12 to +32$^{\circ}$.  Table~\ref{clustercoordinatestable} lists the 2MASS coordinates and key properties of each cluster.  The clusters were originally optically-discovered and span only a modest range of reddening values.  Some of the clusters, such as M 67, also lie well off the Galactic plane.

Archival 2MASS $J$, $H$, and $K_S$ photometric data were chosen to compare to theoretical isochrones for two main reasons: 1) 2MASS provides a consistent photometric dataset for the entire cluster sample, and 2) the all-sky coverage of 2MASS allows for comprehensive measurements of background stellar contamination for every cluster \citep[e.g.,][]{Bonatto2007, Alves2012}.  Using additional datasets, such as the Sloan Digital Sky Survey (SDSS), might reduce the uncertainties of some of the derived parameters for some of the clusters, but such data were not available for all clusters in the sample.  So, using these data for only some of the clusters would have introduced bias.

We obtained 2MASS data from the IPAC Infrared Science Archive (IRSA) consisting of 50 arcmin diameter fields centered on each cluster.  This size was chosen to sufficiently characterize the level of background stars even for clusters spanning large angular radii.  $(J-K)$ versus $H$ CMDs were created from these data and the CMDs were used in fitting theoretical isochrones.  The radius of the selected area was the same for each cluster, regardless of cluster size.

\section{Fitting Theoretical Isochrones to Cluster CMDs} \label{fitting isochrones}
To find cluster properties, along with corresponding uncertainties, we developed a multi-step process to fit theoretical isochrones to cluster CMDs.  These steps are summarized here and discussed in greater detail below.  We used the PAdova and TRieste Stellar Evolution Code (PARSEC) isochrones from the Padova set of models \citep{Bressan2012}. Because the isochrones used are not continuous analytic functions, Monte Carlo realizations of the isochrones were created for the purpose of comparing to cluster CMDs.  

For each cluster, we created two Hess diagrams \citep[e.g.,][]{Alves2012} of the CMD densities of 2MASS stars: one of the region containing the cluster and one of a region used to characterize the background surrounding the cluster.  The background CMD was subtracted from the cluster CMD, and Monte Carlo realizations of the CMD overdensity were created.  Monte Carlo realizations of the theoretical isochrones, following the initial mass function (IMF) of \citet{Chabrier2005},  were also created at discrete points in a parameter space of log(age), distance modulus [(m-M)$_H$], and color excess [E$(J-K)$].  At each point, an isochrone realization was generated at that age and shifted by the (m-M)$_H$ and E$(J-K)$ values.  Each cluster background-subtracted CMD realization was compared to the set of isochrone realizations over the entire parameter space.  A $\chi^2$ statistic was computed at each point in parameter space to determine the goodness-of-fit between each isochrone realization and each cluster CMD.  The parameters of age, distance, and reddening that yielded the minimum $\chi^2$ for each cluster were adopted as best representing the cluster properties.

\subsection{Background Subtraction}\label{background subtraction}
The contamination of cluster CMDs by foreground and background stars is less severe in the outer Galaxy than in the inner Galaxy.  However, for some clusters in the sample, there remains a significant level of contamination, especially for those close to the Galactic disk.  To find accurate cluster properties, it is necessary to remove the effects of contaminating stars.  However, determining cluster membership on a star-by-star basis is difficult when using only photometric information.  Several studies have used photometric properties, sometimes in conjunction with other available cluster data, to select cluster members \citep[e.g.,][]{Currie2010, Maia2010, Dias2012, Alves2012}.  For the purpose of finding the best-fit isochrone for each cluster CMD while minimizing background contamination, we converted each cluster CMD into a CMD image\footnote{Because there were several steps involved in the fitting procedure, a distinct name was given to the output of each step.  We summarize these names here for clarity.  A ``CMD image" (Section \ref{cmd images}) is a Hess diagram, created by binning the color-magnitude space of a cluster into pixels and counting the number of stars that fall into each pixel.  A ``CMD Poisson draw image" is created from a ``CMD image" by drawing a value from a Poisson distribution for each pixel, where the mean of the Poisson distribution is equal to the number of stars in the pixel.  A ``cluster overdensity CMD image" is the result of subtracting an outer region CMD Poisson draw image from a corresponding inner region CMD Poisson draw image (Section \ref{cmd image realizations}).  A ``cluster Monte Carlo realization" is created by populating a cluster overdensity CMD image with discrete stars (Section \ref{creating cluster realizations}).  An ``isochrone realization" is a Monte Carlo realization consisting of individual stars, following an IMF, that represents a given isochrone curve (Section \ref{generating an isochrone realization}).  A ``synthetic cluster" (Section \ref{synthetic clusters}) is a cluster CMD realization generated from a theoretical isochrone at a specified age and metallicity, which is shifted by a given distance modulus and reddening, and whose realization stars were scattered in the CMD by typical photometric noise.} from which a background CMD image could be subtracted.  These steps are detailed below.

\subsubsection{Background Stellar Densities and Cluster Extents}\label{background stellar densities}
Radial density profiles of the cluster star counts were created from the 2MASS photometric data.  To establish the background stellar density and the cluster angular extent, the star counts were annularly binned, where the bins were 50 arcsec wide and centered on the stellar overdensity in each region.  Each cluster radial density profile was fit with a Gaussian plus a uniform distribution \citep[e.g.,][]{Mercer2005}. 

For each cluster, the 50 arcmin diameter region was separated into three annular sub-regions: an inner region, a buffer zone, and an outer region.  We defined the cluster radius to be 1.5 times the Gaussian width\footnote{The Gaussian width is equal to one Gaussian $\sigma$, but as we use the term $\sigma$ for the parameter uncertainties, the term Gaussian width is used here to reduce confusion.} (i.e., R$_{1.5GW}$) and set the inner region radius to this R$_{1.5GW}$.  The outer region was used to estimate the level of background contamination of field stars in the inner region, and was defined as the area outside of 3.5 times the Gaussian width.  The number of cluster stars, found by integrating the radial density profile, the background contamination level, and the R$_{1.5GW}$ radius of each cluster are listed in Table~\ref{clustercoordinatestable}.  

The buffer zone was an annular gap, from 1.5--3.5 Gaussian widths, located between the inner and outer regions.  Stars located in this region were not used in the analysis.  The use of the buffer zone was necessary to avoid regions where neither cluster members nor field stars dominate significantly, which would make it difficult to separate the two groups.

\begin{figure}
	\begin{center}
		\includegraphics[width = 0.45\textwidth]{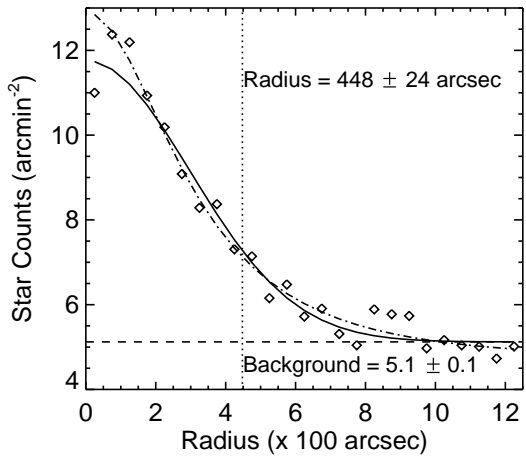}
		\caption{Radial density profile of the stellar counts of NGC 2099 based on 2MASS photometric data, with a Gaussian fit overlaid as the solid black curve.  The cluster radius of $\sim$440 arcsec, defined as 1.5$\times$ Gaussian width, is indicated by the dotted vertical line.  The 3.5$\times$Gaussian width location is outside the plot range.  The mean background star count level of $\sim$5 counts~arcmin$^{-2}$ is indicated by the black dashed horizontal line.  A King profile was also fit to the profile, shown as the dashed-dotted curve.}
		\label{radprofiledensityfig}
	\end{center}
\end{figure}

Figure \ref{radprofiledensityfig} shows the radial density profile for cluster NGC 2099, with its R$_{1.5GW}$ cluster radius indicated.  For a rich, relatively nearby cluster, such as NGC 2099, the cluster angular extent is large.  As can be seen from the figure, enlarging the cluster radius beyond R$_{1.5GW}$ would likely introduce more background stars than cluster stars and increase the contamination of the CMD.  

A King profile, following Eq. 14 of \citet{King1962}, which had the same number of terms as the Gaussian fit, was also fit to the density profile, as shown by the dashed-dotted curve in Figure \ref{radprofiledensityfig}.  While it appears to fit the density profile well, the uncertainties returned (r$_{core}$ = 330 $\pm$ 70 arcsec) are larger than the uncertainties from the Gaussian fit (R$_{1.5GW}$ = 448 $\pm$ 24 arcsec).  In addition, it was not clear that a King profile was the appropriate function to fit to the sample, as some of the clusters are young [log(age)$\sim$7] and may not be dynamically relaxed.  For these reasons, only Gaussian distributions were fit to the cluster radial density profiles.  

\subsubsection{CMD Images}\label{cmd images}

For each cluster, separate $(J-K)$ vs $H$ CMDs were created for the inner and outer regions.  For this process, and for the subsequent processes of creating cluster realizations and fitting to theoretical isochrones, we only used stars that were brighter than 15$^{th}$ mag in $H$-band because the 2MASS photometric uncertainties increase significantly beyond this threshold.  

\begin{figure}
	\begin{center}
		\includegraphics[width = 0.45\textwidth]{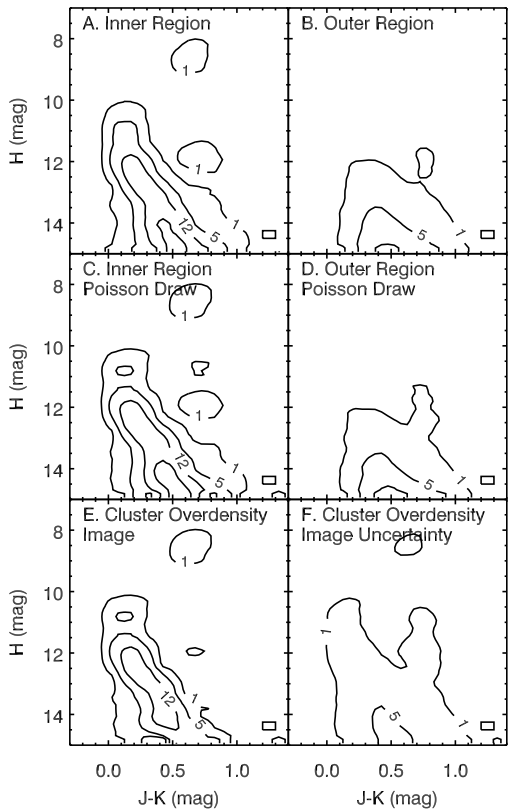}
		\caption{Illustration of the steps used to create an overdensity CMD image for NGC 2099.  Top panels (A and B) show the inner and outer region CMD images, while the middle panels (C and D) show Poisson draws from the CMD images of A and B.  The bottom left panel (E) shows the net overdensity of the star counts after the counts of D have been subtracted from those of C.  Panel F shows the uncertainty of the overdensity image of Panel E.  The size of each pixel in color-magnitude space is shown as a reference rectangle in the bottom right of each panel.  The contour levels are drawn at [1, 5, 12, and 20] counts per pixel.}
		\label{clusgencartoonfig}
	\end{center}
\end{figure}

The individual stars in each CMD were binned into pixels that were 0.1 mag wide in color and 0.25 mag wide in apparent magnitude to create stellar density CMD images (Hess diagrams) that were 26$\times$36 pixels.  These pixel sizes were chosen to retain sensitivity to small variations in the star counts across the CMD, but also to be large enough to have significant star counts.  The count value for each pixel was equal to the number of stars that fell into the pixel color and magnitude ranges.  The outer region CMD counts were scaled by the ratio of the areas of the inner region to the outer region.  Figure \ref{clusgencartoonfig} (Top, A and B) shows the inner and outer region CMD images for NGC 2099.  The inner CMD image (Panel A) shows a well-defined main sequence and a red clump, whereas the outer region image (Panel B) shows no cluster-like features.  We conclude that the outer region contains very few cluster members.  

\subsubsection{CMD Poisson Draw Images}\label{cmd image realizations}
To create the cluster CMD image corrected for background contamination, the simplest approach would be to directly subtract the area-scaled outer region CMD image from that of the inner region.  However, doing so would yield incorrect uncertainties because stellar counts are governed by Poisson statistics in both regions.  Instead, a statistical approach was developed that preserved the Poisson nature of the stars in all regions.  

A schematic of this approach continues in Figure \ref{clusgencartoonfig} for NGC 2099.  Panels A and B of Figure \ref{clusgencartoonfig} show the inner and outer region CMD images.  At every pixel in each of the CMD images, a value was drawn from a Poisson distribution whose mean equaled the count in that pixel\footnote{For the outer region, a Poisson draw was created of the CMD image before the counts were scaled by the ratio of the areas between the inner and outer regions.  This Poisson draw was then scaled by the ratio of the areas.}.  Panels C and D show the resulting CMD Poisson draw images of the inner and outer region CMD images, respectively, for a single Poisson draw.  These images closely resemble the CMD images of Panels A and B, but because they are draws from Poisson distributions, the counts of the pixels will fluctuate with each draw.

The outer region CMD Poisson draw image was subtracted from the inner region CMD draw to create a cluster overdensity CMD image.  Panel E of Figure \ref{clusgencartoonfig} shows a cluster overdensity image for NGC 2099, which was produced by subtracting the Panel D outer region draw from the Panel C inner region draw.  Panel F shows the Panel E image uncertainty, propagated from the Poisson uncertainties of Panels C and D.  A set of such cluster overdensity images for any one cluster will exhibit pixel count fluctuations governed by Poisson statistics that reflect the original inner and outer region properties.  For sparse clusters, these overdensity images will exhibit more fluctuations because the lower cluster star counts will be more sensitive to background Poisson noise.

NGC 2099 is a well-defined cluster, and therefore, is an easy case on which to perform the process illustrated in Figure \ref{clusgencartoonfig}.  We generated the same figure for King 1 (not shown), which is a less populated cluster.  The same features seen in Figure \ref{clusgencartoonfig} for NGC 2099 are similarly seen for King 1, such as the prominence of the main sequence in the overdensity image, which indicates that the procedure is robust.

\subsection{Creating Cluster Monte Carlo Realizations}\label{creating cluster realizations}
The method developed to match isochrones to cluster CMDs, described below, compares the colors and magnitudes of individual stars.  Therefore, we converted each cluster overdensity CMD image into a cluster CMD Monte Carlo realization of individual points representing stars.   Each pixel of the overdensity CMD image was populated with a number of realization stars equal to the star count value in that pixel.  The assigned colors and magnitudes of the realization stars in each pixel followed a uniform distribution within the bounds of the color and magnitude ranges of the pixel.  Each realization star was assigned color and magnitude uncertainties based on the average 2MASS photometric uncertainties at that color and magnitude.  These color and magnitude uncertainties were computed from the 2MASS photometric data of a randomly chosen 2.25 square degree field of view.

To prevent outliers in the CMD Monte Carlo realization from adversely influencing the isochrone fit, stars far from the CMD main sequence and red clump were removed before comparing to isochrones.   The Monte Carlo realization was temporarily converted to an image, similar to the process described in Section \ref{background subtraction}, by binning the stars into color-magnitude pixels 0.1 mag wide in color and 0.25 mag wide in apparent magnitude.  At each 0.25 mag step in magnitude, the number of stars per pixel along the color axis was fit with a Gaussian function.  Individual realization stars that were located at colors further than three times the Gaussian width from the Gaussian peak were removed from the realization.  No stars were eliminated when the Gaussian width was larger than 1 mag in color, as these magnitude bins were too sparsely populated to generate a significantly peaked overdensity.

\begin{figure}
	\begin{center}
		\includegraphics[width = 0.45\textwidth]{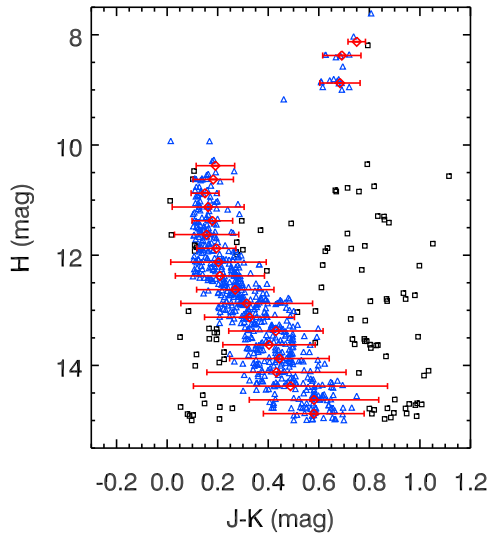}
		\caption{A Monte Carlo realization of an overdensity CMD image of NGC 2099.  Triangles and squares represent realization stars.  Centers of the Gaussian profiles of each magnitude step are indicated by red diamonds, and the [-3$\sigma$, +3$\sigma$] width intervals of the steps are shown by the red horizontal error bars.  Blue triangles represent realization stars that fall within three times the Gaussian widths.  Black squares represent stars that fall outside this range, which were not included in the subsequent isochrone fitting.}
		\label{cmdrealslicefig}
	\end{center}
\end{figure}

Figure \ref{cmdrealslicefig} shows a CMD Monte Carlo realization for NGC 2099, which was fit by Gaussian functions of star counts along the color axis at each 0.25 mag step.  Note that the actual binned pixel counts are not shown in the figure.  The red clump and main sequence are well defined.  Stars that are located further than three times the Gaussian width from the main sequence were flagged as outside the acceptable range and rejected from use (open black squares in the figure).  The resulting ``trimmed'' cluster Monte Carlo realizations were used in the following steps.

The number of stars populating the cluster Monte Carlo realizations fluctuated based on the Poisson draws of the inner and outer region CMD images.  The exact color and magnitude locations of the realization stars within each color-magnitude pixel also fluctuated with each draw.  Therefore, in the following procedure, we generated multiple cluster CMD realizations to capture the full ranges of these fluctuations.

\subsection{Isochrone Fitting Procedure} 
We developed a $\chi^2$ minimization approach to find the PARSEC isochrone \citep{Bressan2012}, at a given age and metallicity and shifted by distance modulus and color excess, that best represented each cluster to find the cluster properties of age, distance, and reddening.  Monte Carlo realizations of model stars were created from the isochrones.  The cluster CMD Monte Carlo realizations, described above, were then compared to these isochrone realizations, as described below.

\subsubsection{Generating an Isochrone Realization}\label{generating an isochrone realization}
Monte Carlo realizations of each isochrone were created following an IMF, where each isochrone was populated with individual points representing stars \citep[e.g.,][]{Janes2013}.  These isochrone realizations were generated for every cluster, where the number of isochrone realization stars was set equal to the number of cluster CMD realization stars.  We adopted a ``kinked" IMF based on Eq. 1 in \citet{Chabrier2005} :

\begin{equation}\label{equation imf}
\begin{split}
 \xi(log~(m)) = \frac{dn}{dlog~(m)} = 0.093 ~~~~~~~~~~~~~~~~~~~~~~~~~& \\
\times exp\left\{-\frac{(log~(m) - log~(0.2))^2}{0.605}\right\}, \\
 m_{min} \le m \le 1 M_{\odot} ~~~~~~~~~~~~\\
  = 0.041~m^{-1.35\pm 0.3}~ ,m_{max} \ge m\ge 1 M_{\odot}
 \end{split}
\end{equation}

The mass range of each isochrone was uniformly divided into 50 bins, where the mass limits were set to the minimum (m$_{min}$) and maximum  (m$_{max}$) masses of the isochrone.  These mass ranges were slightly different for every isochrone.  Each mass bin was populated by the relative number density of stars per mass, based on Eq. \ref{equation imf}.  

\begin{figure}
	\begin{center}
		\includegraphics[width = 0.45\textwidth]{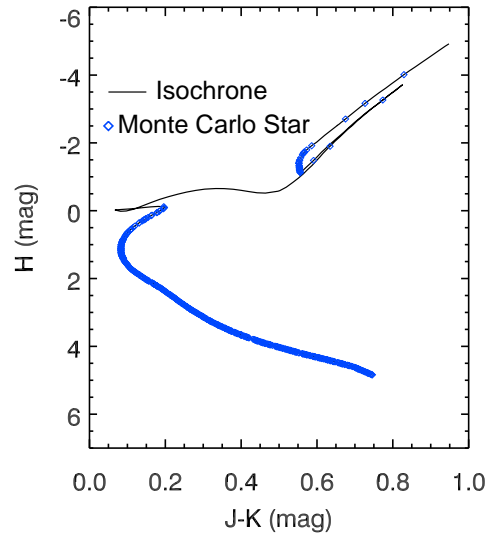}
		\caption{An example of a Monte Carlo realization of an isochrone.  The solid black line shows the PARSEC isochrone at an age of 1 Gyr and Z = 0.019.  The blue diamonds show 600 Monte Carlo realization stars that follow the \citet{Chabrier2005} IMF.}
		\label{isorealfig}
	\end{center}
\end{figure}

Figure \ref{isorealfig} shows a 1 Gyr, solar metallicity $(J-K)$ versus $H$ PARSEC isochrone with a Monte Carlo realization overlaid.  The realization contains 600 stars, shown as open blue diamonds, which are distributed following the \citet{Chabrier2005} IMF.  The isochrone realization reveals the expected concentration of stars at the theoretical red clump location and along the main sequence.  

\subsubsection{Finding the Best-Fitting Isochrone}\label{determining the best-fitting isochrone}
To find the best-fit isochrone for each cluster, we tested isochrones across a 3D parameter grid of log(age), distance modulus [(m-M)$_H$] (uncorrected for extinction), and color excess [E$(J-K)$].  At each grid point, an isochrone was selected at that age and shifted in magnitude by the distance modulus and shifted in color by the color excess. 

We tested whether having metallicity as another free parameter affected the results of the fitting procedure, and found that, in most cases, it did not (see Section \ref{testingmetallicity}).  Based on these results, metallicity was not fit for the majority of the clusters, but was instead fixed for each cluster.  If metallicity had been determined spectroscopically for a cluster in a previous study, that metallicity was adopted (see Table~\ref{clustercoordinatestable}).  Otherwise, solar metallicty was assumed.  The exception was cluster NGC 1857, for which a metallicity of Z = 0.005 was used.  In this case, solar metallicity isochrones would not fit both the main sequence and the small apparent red clump.  We attempted to fit isochrones using several different metallicities, and determined that a metallicity of Z = 0.005 provided the best fit to the cluster CMD.  

The isochrone metallicity used for each cluster is listed in Table \ref{clustercoordinatestable}.  While the PARSEC isochrones adopt a value of Z = 0.0152 for solar metallicity, to remain consistent with earlier studies, we adopted a metallicity of Z = 0.019 \citep[e.g.,][]{Anders1989, Girardi2000} as the solar metallicity.  Therefore, in the context of the PARSEC models, we have selected a solar metallicity that is somewhat metal-rich.

The 3D parameter grid consisted of 20 steps in each direction of log(age), distance modulus, and color excess.  We performed the isochrone fitting procedure, described below, twice: once for a coarsely stepped grid of parameters and a second time for a more finely stepped grid.  For the coarse grid, log(age) was stepped by 0.05, distance modulus by 0.075 mag, and color excess by 0.025 mag.  The initial center of each cluster's coarse grid in parameter space was determined by visually identifying the isochrone that appeared to best match the cluster CMD \citep[e.g.,][]{Alves2012}.  The finer grid, centered on the best-fitting point found from the coarse grid, consisted of log(age) steps of 0.025, distance modulus steps of 0.0375 mag, and color excess steps of 0.0125 mag.

A cluster CMD Monte Carlo realization, generated following the steps in Section \ref{creating cluster realizations}, and an isochrone realization, following Section \ref{generating an isochrone realization} with the same number of stars as the cluster realization, were compared at each parameter grid point.  One of the key goals of the fitting procedure was to give more weight to stars with lower photometric uncertainties, i.e., brighter stars.  Therefore, cluster CMD realization stars were ranked by brightness.  A scaled distance ($D_{S_I}$)  was calculated between the brightest cluster star and each isochrone realization star in CMD space, defined as:

\begin{equation}\label{equation lower chisq}
D_{S_I} = [(\frac{(J-K)_C - (J-K)_I}{\sigma_{{(J-K)}_C}})^2 + (\frac{H_C - H_I}{\sigma_{H_C}})^2)]^{1/2} , 
\end{equation}
where $C$ refers to the cluster realization star and $I$ refers to one isochrone realization star.  $(J-K)_C$ and $(J-K)_I$ are the colors of the cluster star and the isochrone star, respectively, $H_C$ and $H_I$ are the $H$ magnitudes of the cluster star and isochrone star, respectively, and $\sigma_{{(J-K)}_C}$ and $\sigma_{H_C}$ are the uncertainties of the $(J-K)$ color and $H$ magnitude of the cluster star, respectively.  Because the cluster stars came from Monte Carlo realizations, they do not have intrinsic uncertainties.  So, the uncertainties assigned to them were based on the average 2MASS uncertainties at the same $J$, $H$, and $K$ magnitudes as the cluster stars.  

The isochrone star that yielded the lowest $D_{S_I}$ value was paired to the cluster star and removed from the pool.  Another set of D$_{S_I}$ values was computed using the next brightest cluster realization star and the remaining isochrone realization stars, and the isochrone star that yielded the lowest D$_{S_I}$ value was paired to the cluster star.  This process was continued until every cluster star was paired to one isochrone star. 

\begin{figure}
	\begin{center}
		\includegraphics[width = 0.45\textwidth]{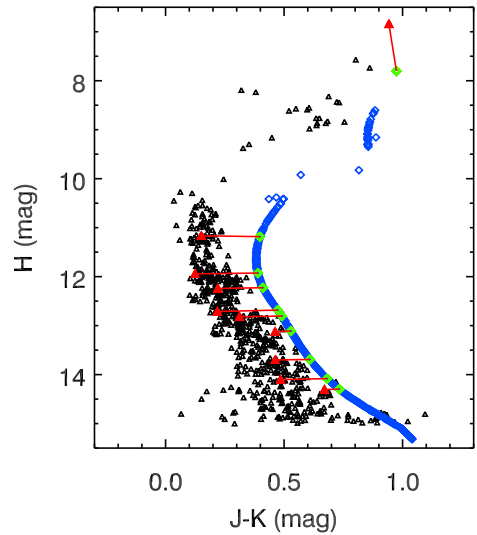}
		\caption{Example of the process of matching the cluster CMD Monte Carlo realization stars to the isochrone realization stars to determine the best-fit isochrone for NGC 2099.  The $\sim$630 cluster realization stars are shown as black triangles and the $\sim$630 isochrone stars are shown as blue diamonds.  Ten cluster stars (every 60th star, ranked by brightness) are shown connected to their paired isochrone stars with red lines, where the paired cluster and isochrone stars are displayed as red triangles and green diamonds, respectively.  Note that this example represents a poor match between the isochrone and cluster CMD, and was chosen to better display the pairing process.}
		\label{matchstarsfig}
	\end{center}
\end{figure}

A representation of this process is shown in Figure \ref{matchstarsfig}.  Select cluster realization-isochrone realization stellar pairs are highlighted.  The brightest cluster realization star, at $\sim$7 mag in $H$-band, is paired to the isochrone star that yielded the lowest $D_{S_I}$ value.  Each cluster realization star was paired with an isochrone star, but only 10 cluster stars, every 60th in the brightest ranked star list, are highlighted in Figure \ref{matchstarsfig}.  Allowing the brightest clusters stars to be matched first ensured that these stars had somewhat greater influence on the fit.  

Once every cluster-isochrone stellar pair was created, the $D_{S_I}$ values of every pair were squared and summed to create a $\chi^2$ statistic to represent the goodness-of-fit between the cluster Monte Carlo realization and isochrone Monte Carlo realization:

\begin{equation}\label{equation upper chisq}
\chi^2 = \sum\limits_{p} D_{S_p}^2, 
\end{equation}
where p represents each cluster-isochrone stellar pair.    

For every cluster, 30 sets of Poisson draws were created from the inner and outer region CMD images, from which 30 CMD overdensity images were created.  One Monte Caro realization was created from each of the 30 overdensity images, resulting in thirty cluster CMD Monte Carlo realizations for each cluster.  Ten Monte Carlo isochrone realizations were created at each point in the 3D parameter grid.  Therefore, at each point in the parameter grid, to assess a particular isochrone's goodness-of-fit, 300 $\chi^2$ values were computed via Eqs. \ref{equation lower chisq} and \ref{equation upper chisq}.  These 300 $\chi^2$ values were median-filtered and a mean $\chi^2$ was computed to represent the goodness-of-fit at each grid point.  This is a more conservative approach than simply selecting the smallest $\chi^2$ of the 300 values, which could easily be an outlier.

The point in the 3D parameter grid that had the lowest mean $\chi^2$ was chosen as the point whose isochrone parameters of log(age), distance modulus, and color excess best represented the actual cluster properties.

The uncertainties of the best-fit parameters were estimated by evaluating the mean $\chi^2$ values of the grid.  To estimate the uncertainty in log(age), the distance modulus and color excess were held constant at their best-fit points, and the mean $\chi^2$ values of the 20 points along the log(age) axis of the parameter grid were selected.  The uncertainty in log(age) was found by calculating the $\chi^2$ weighted deviation of these 20 log(age) values:

\begin{equation}
\sigma_{x_{min}} = \sqrt{\frac{\sum((x_i - x_{min})^2/\chi^2_i)}{\sum(1/\chi^2_i)}},
\end{equation}
where the $x_i$ are the 20 log(age) values along the log(age) parameter axis and the $\chi^2_i$ represent the 20 corresponding $\chi^2$ values.  The best-fit value of each parameter is then $x_{min} \pm \sigma_{x_{min}}$, where $x$ is either log(age), distance modulus, or color excess.

\begin{figure}  
	\begin{center}
		\includegraphics[width = 0.45\textwidth]{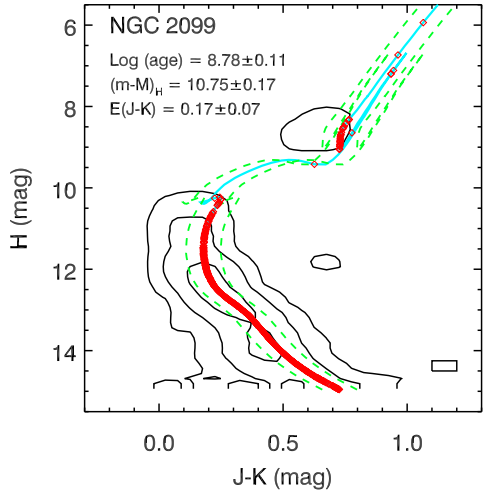}
		\caption{NGC 2099 overdensity CMD image with the best-fit isochrone overlaid.  Cluster overdensity contours at levels of [1, 5, and 12] counts per pixel are shown in black, and the best-fit isochrone is shown as the solid blue line.  A Monte Carlo realization of the best-fit isochrone is overlaid, with $\sim$630 stars shown as red diamonds.  The best-fit parameters are listed, along with their corresponding uncertainties.  The best-fit isochrone does not fall directly in the center of the red clump (see text).  This isochrone was shifted by $\pm$1$\sigma$ in E$(J-K)$ and drawn as the dashed green lines.  The size of each pixel in the CMD image in color-magnitude space is shown as a reference rectangle in the bottom right of the panel.}
		\label{ngc2099bestfitfig}
	\end{center}		
\end{figure}

One cluster overdensity CMD image for NGC 2099 is shown in Figure \ref{ngc2099bestfitfig}, with the best-fit isochrone overlaid.  This isochrone does not appear to fit the red clump perfectly, but it does fall directly down the middle of the main sequence.  This discrepancy may be due to the relatively small number of stars  in the red clump compared to the large number along the main sequence, which dominates the fit.  As can be seen by the $\pm$1$\sigma$ $E(J-K)$ isochrones overlaid on the plot (dashed green lines), the $-1\sigma$ limit isochrone that falls in the middle of the red clump does not fit the main sequence as well.  This result indicates that the isochrone selected does best represent the overall shape of the cluster CMD at the given spectroscopic metallicity.  We also tested whether using different metallicities would affect the fit, and those results are described in Section~\ref{testingmetallicity}.

\subsection{Results of Isochrone Fitting}\label{results of isochrone fitting}
Of the 31 clusters in the NIR polarization-based sample, 24 were fit successfully.  We determined whether a cluster was successfully fit by a visual comparison between the best-fit isochrone and the cluster CMD.  For the seven that were not fit successfully, the best-fit isochrones failed to overlap the cluster main sequence and/or the red clump of the CMDs.  These seven clusters, Berkeley 14, Berkeley 18, Berkeley 70, Basel llb, Berkeley 32, NGC 2126, and Berkeley 39, were either faint in the NIR or sparse, thereby limiting the success of the fitting procedure.

\begin{figure*} 
	\begin{center}
		\includegraphics[width = 0.8\textwidth, angle=90]{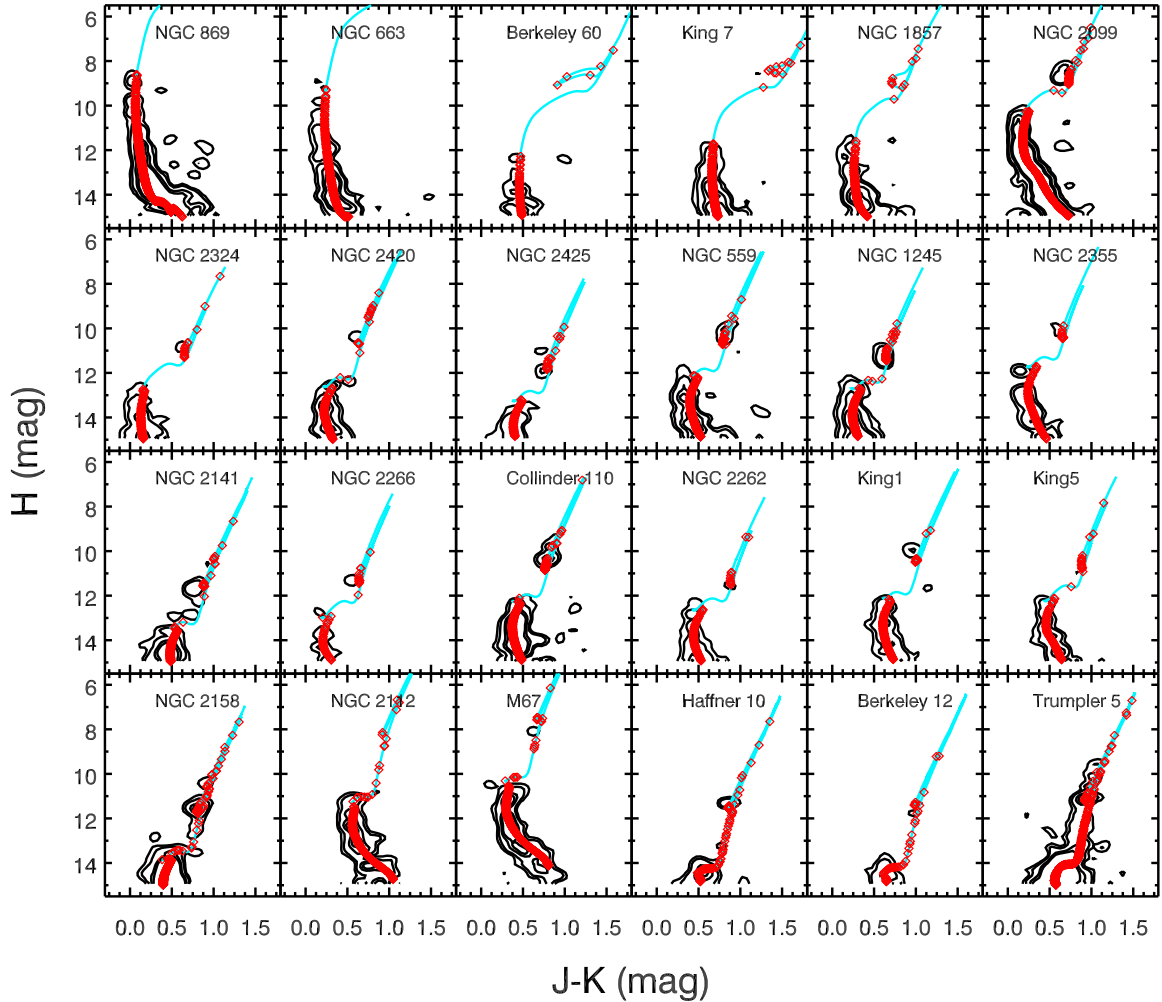}
		\caption{Similar to Figure \ref{ngc2099bestfitfig}, but for the 24 clusters for which all three parameters of log(age), distance modulus, and color excess were fit.  Contour levels are shown at [1, 2, 5, and 7] stars counts per pixel.  The clusters are plotted in order of increasing age to show the evolution of the shape of the isochrone as well as the distribution of stars in the CMD.}
		\label{allcmdbestfitfig}
	\end{center}
\end{figure*}

Figure \ref{allcmdbestfitfig} shows best-fit isochrones for the 24 successfully-fit clusters. While the clusters are listed in order of Galactic longitude in the tables, they are ordered by age in Figure~\ref{allcmdbestfitfig} to show the progression of the shape of the cluster CMDs and the distribution of stars as a function of age.  As clusters age, the red clump becomes more populated.  As they age further, the asymtotic giant branch (AGB) becomes more populated as more stars evolve off the main sequence.  This evolution of the loci of stars in cluster CMDs can be seen in Figure~\ref{allcmdbestfitfig}.  The youngest cluster, NGC 869 has no observed red clump, whereas the intermediate age NGC 2099 contains a red clump.  In the oldest cluster, Trumpler 5, the evolved stars are located along the AGB.

\begin{figure*} 
 	\begin{center}
 		\includegraphics[width=0.8\textwidth, angle =90]{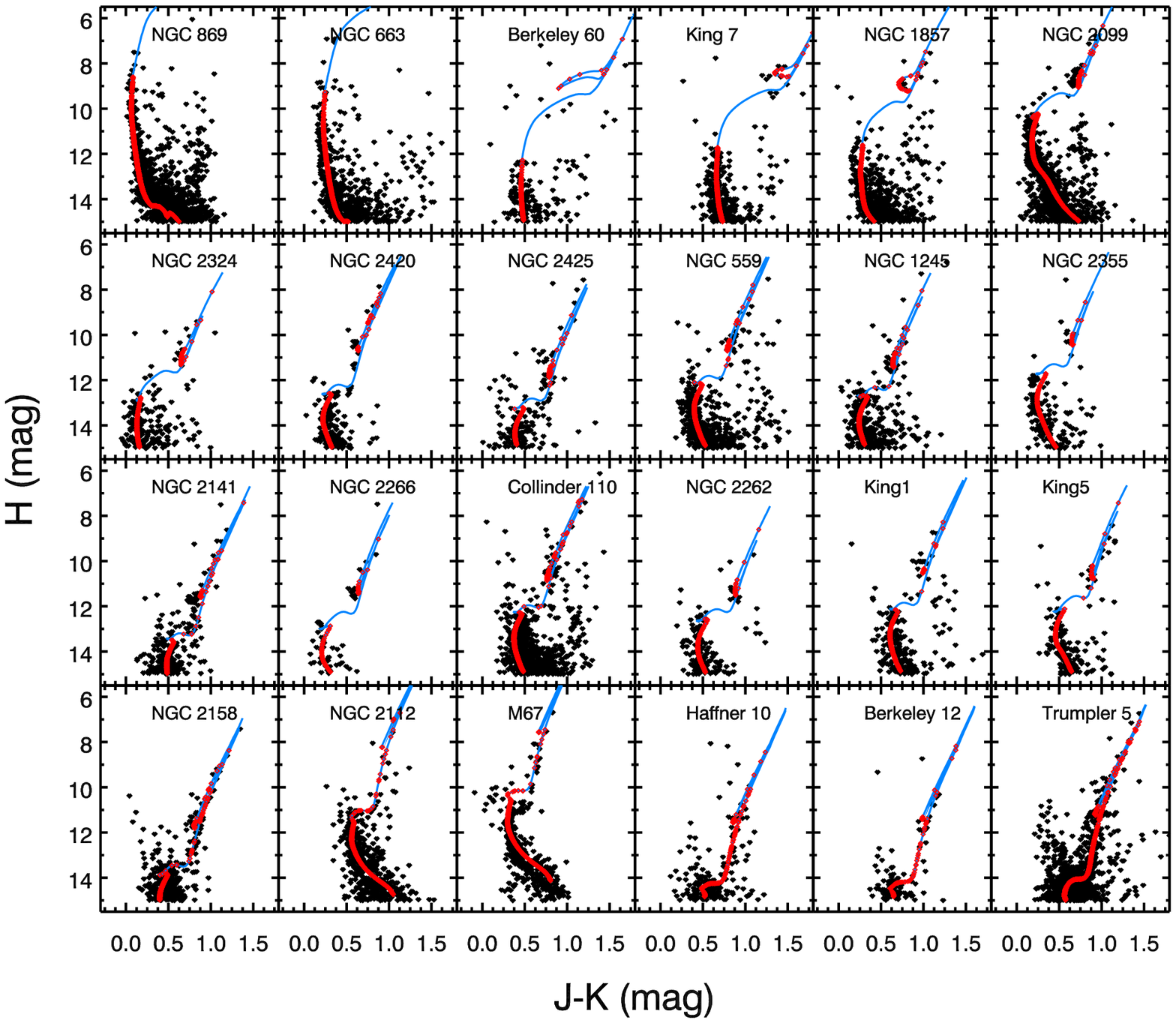}
 		\caption{Similar to Figure \ref{allcmdbestfitfig}, but with the individual 2MASS $(J-K)$ and $H$ colors and magnitudes of all stars located in each cluster inner region plotted as black diamonds.  The best-fit isochrones trace the cluster features for the 24 clusters.}
 		\label{allcmdstarsbestfitfig}
 	\end{center}
 \end{figure*}

Figure \ref{allcmdstarsbestfitfig} is similar to Figure \ref{allcmdbestfitfig}, but instead of plotting the best-fit isochrones on the cluster overdensity CMD images, the isochrones are plotted on CMDs of the individual 2MASS stars from the inner region of each cluster.  Because all stars located in each inner region are included, the contamination from background stars is present.  Nevertheless, the best-fit isochrones trace the loci of cluster stars.

\begin{figure}  
	\includegraphics[width=0.45\textwidth]{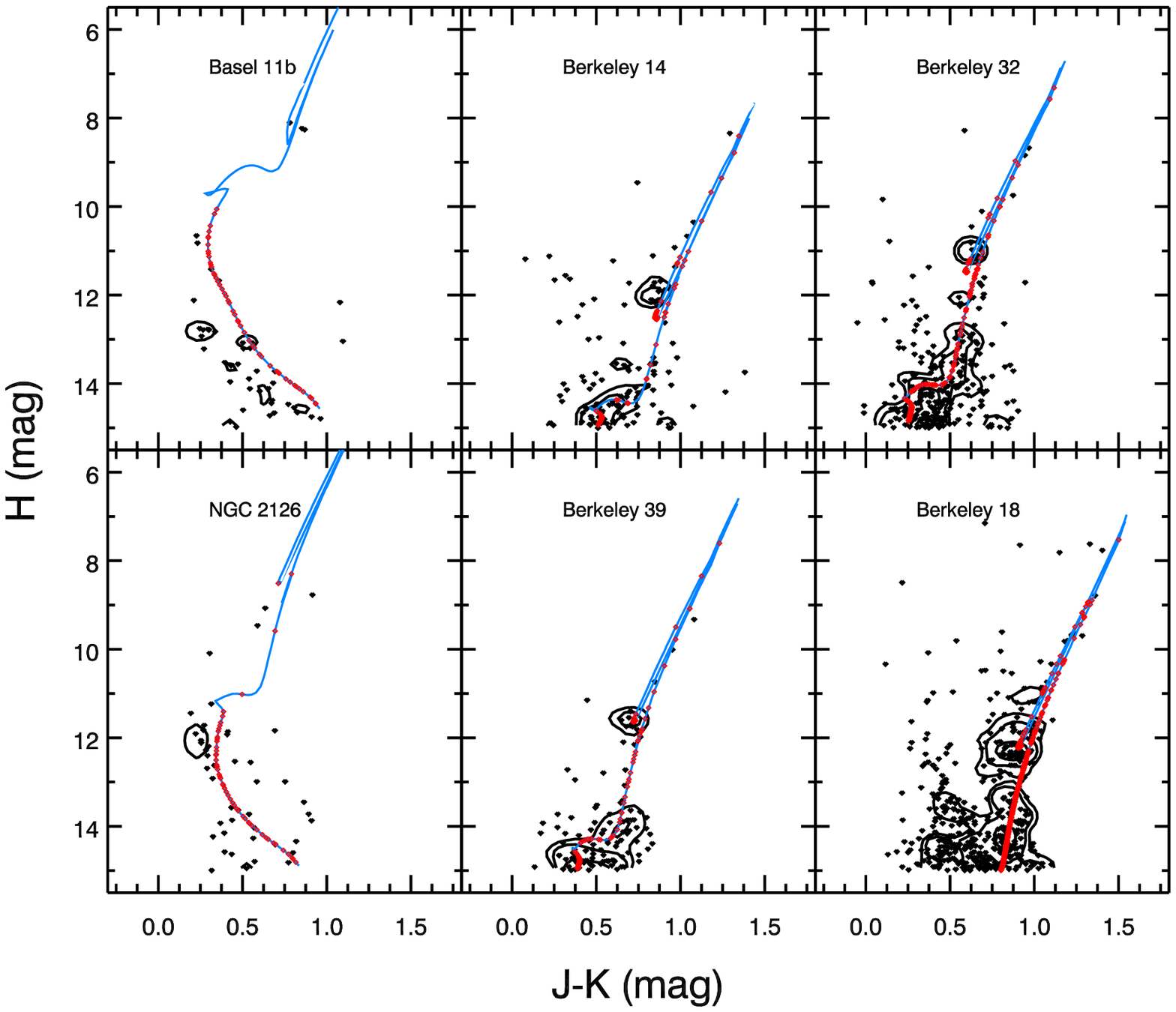}
	\caption{Similar to Figures \ref{allcmdbestfitfig} and \ref{allcmdstarsbestfitfig} but for the six clusters where log(age) was fixed and only distance modulus and color excess were fit.  The black contour levels at [1, 2, 5, and 7] counts per pixel show the cluster overdensity CMD images, and the black diamonds are individual 2MASS stars located in the cluster inner regions.}
	\label{rejcmdbestfitfig}
\end{figure}

For the seven clusters that were not fit successfully for all three parameters, we repeated the procedure detailed above, but kept ages fixed, and only searched the 2D parameter space of distance modulus and color excess.  The fixed age of each cluster was adopted from the most recent published study to derive the properties \citep{Kharchenko2013} of these clusters.  By only fitting for two parameters, we recovered the distance modulus and color excess for six of the seven clusters.  Berkeley 70 was still not properly fit, as it is too faint.  Figure \ref{rejcmdbestfitfig} shows the best-fit isochrones overlaid on the six cluster overdensity CMD images and the 2MASS stars from the inner regions, similar to Figures \ref{allcmdbestfitfig} and \ref{allcmdstarsbestfitfig}.
 
Table \ref{clusterbestfittable} lists the best-fit parameters of log(age), (m-M)$_H$, and E$(J-K)$, along with their corresponding uncertainties for the 30 clusters that were fit.  The distances (in pc) and E$(B-V)$, derived from the (m-M)$_H$ and E$(J-K)$, are also listed.  E$(B-V)$ was calculated by equating $A_V = R_VE(B-V)$ and $A_V = rE(J-K)$, where $E(B-V) = rE(J-K)/R_V \approx~E(J-K)/0.53$.  $R_V$ and $r$ were assumed to be 3.1 and $\sim$5.9, respectively, appropriate for diffuse regions in the ISM \citep[e.g.,][]{Whittet2001}.

The approximate numbers of stars in the cluster CMD Monte Carlo realizations are also listed in Table \ref{clusterbestfittable}.  The numbers of CMD stars were estimated by subtracting the total (area-scaled) star counts of each outer region CMD image from that of its inner region CMD image, which was essentially the total number of counts in the cluster CMD overdensity image.  The uncertainties of the number of stars were propagated from the star counts of the CMD images of the inner and outer regions.  The number of stars in each CMD Monte Carlo realization will fluctuate about this number based on the Poisson draw from each cluster CMD overdensity image.  These Table \ref{clusterbestfittable} numbers are less than the number of cluster stars estimated by integrating over the cluster radial density profiles (listed in Table~\ref{clustercoordinatestable}).  This is because Table~\ref{clusterbestfittable} only accounts for stars within the cluster inner regions (R$_{1.5GW}$) and does not include stars fainter than 15$^{th}$ mag in $H$-band.

\subsubsection{Testing Isochrone Fits For Metallicity Dependence}\label{testingmetallicity}
To determine whether adding metallicity as a free parameter would result in improved isochrone fits to the cluster CMDs, we repeated a part of the fitting procedure for five different metallicities for five of the clusters.  We fit the fine grid of parameters at metallicities of Z = 0.005, 0.008, 0.012, 0.017, and 0.019.  The five clusters selected were King 1, NGC 1245, King 7, NGC 2099, and NGC 2141.  These were chosen because they had a variety of ages, and had a combination of spectroscopically determined (NGC 1245, NGC 2099, NGC 2141, as listed in Table 1) and assumed (King 1, King 7) metallicities as noted in Table~\ref{clustercoordinatestable}.

For each of the five clusters, the resulting five minimum $\chi^2$ values of the best-fit locations in the parameter grids, corresponding to the five metallicities, either remained unchanged or increased at low metallicities compared to the minimum $\chi^2$ value of the metallicity initially used in the isochrone fits, listed in Table~\ref{clustercoordinatestable}.  For all five clusters, the three best-fit parameters found at each metallicity were within 1$\sigma$ of the best-fit parameters listed in Table 2.  Therefore, varying the metallicity did not change the best-fit parameters by more than 1$\sigma$.  Visual inspection of the best-fit isochrones overlaid on the cluster CMDs also showed no improvement, with the exception of NGC 2099.  At a metallicity of Z = 0.008, the red clump of the best-fitting isochrone was shifted blueward by about 0.03 mag in $(J-K)$ compared to the red clump location of the Z = 0.019 best-fit isochrone (shown in Figure \ref{ngc2099bestfitfig}).  While this shift places the isochrone red clump closer to the center of the cluster red clump, the amount of the shift is well within the fitting uncertainties of the color excess and is not statistically significant.  Therefore, the best-fit parameters found with Z = 0.019 for NGC 2099 are reported in Table \ref{clusterbestfittable}.  We conclude that the NIR colors are not ideal for determining metallicity, as optical colors are more sensitive to metallicity changes.

\subsection{Quality Testing using Synthetic Clusters}\label{synthetic clusters}
To determine the accuracy of the fitting procedure, we created and fit synthetic clusters using the same procedure as described above.  The synthetic clusters were generated from isochrone realizations at 12 ages, ranging from log(age) of 7.3 to 9.6, created using the procedure described in Section \ref{generating an isochrone realization}.  The $(J-K)$ colors of the stars were shifted by 0.25 mag, and the distance modulus (uncorrected for extinction) by 10 mag, to simulate clusters with reddenings of E$(B-V)$ of 0.47 mag at distances of $\sim$900 pc.  The stars were then distributed about their color and magnitude values based on the average 2MASS uncertainties corresponding to those colors and magnitudes.  Clusters consisting of such ``synthetic" stars at the 12 given ages were generated with each of 50, 100, and 200 cluster members.

Background field contamination was created using 2MASS photometric data from a 2.25 sq. degree field.  Similar to the procedure described in Section \ref{cmd images}, an outer region CMD image was created from this field.  Each synthetic cluster was assigned a ``cluster'' radius of 160 arcsec, the average of the cluster radii listed in Table \ref{clustercoordinatestable}, to scale the counts of the outer region CMD image.

The synthetic clusters consisting of individual stars were converted to CMD images, and then added to Poisson draws of the outer region CMD image to simulate inner region CMD images (a cluster with field contamination).  Poisson draws were created from the inner and outer region CMD images.  The Poisson draws of the outer region images were subtracted from those of the inner region to create CMD overdensity images.  Monte Carlo realizations of the CMD overdensity images were compared to Monte Carlo realizations of isochrones in 3D parameter grids of log(age), distance modulus, and color excess.  Thirty Poisson draws of the inner and outer regions were created, from which 30 overdensity images were created. One Monte Carlo realization was created from each overdensity image, resulting in 30 Monte Carlo realizations for each cluster.  Ten isochrone realizations were created at each grid point.  The parameters of the grid point that yielded the lowest mean $\chi^2$ value were adopted as most representative of the cluster properties.

In total, 36 synthetic clusters were fit, consisting of three different numbers of members at 12 different ages, all with constant distance and reddening.

\begin{figure} 
	\begin{center}
		\includegraphics[width=0.45\textwidth]{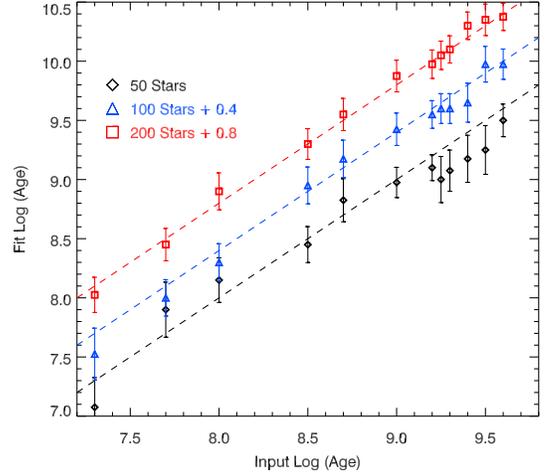}
		\caption{Input log(age) values of synthetic clusters are plotted against output best-fit ages for clusters with 50 (black diamonds), 100 (blue triangles), and 200 (red squares) member stars.  The black, blue, and red dashed lines show where input log(age) = fit log(age) for 50, 100, and 200-star clusters, respectively.  The 100-star and 200-star cluster dashed lines have been displaced vertically by 0.4 and 0.8, respectively, for clarity.  Most of the fit ages are equal to the input age within the fit uncertainties.  Deviations are largest for the 50-star clusters, especially at the older ages.}
		\label{synclusteragefig}
	\end{center}
\end{figure}

Figure \ref{synclusteragefig} shows the resulting best-fit age as a function of the synthetic cluster input age, where the dashed lines represent offset lines of equality.  Most of the fit ages ($>$88\%) fall within their 1$\sigma$ uncertainties of the input ages.  The largest deviations from the line of equality are seen in the clusters with 50 stars.  This result is reasonable, as these clusters have the least number of stars with which to define their main sequences.  They are also the most sensitive to Poisson fluctuations and background contamination.  The points deviate the most around log(ages) of 9.2~--~9.5 as the AGB is not well populated for clusters that have so few stars.

\begin{figure} 
	\begin{center}
		\includegraphics[width = 0.45\textwidth]{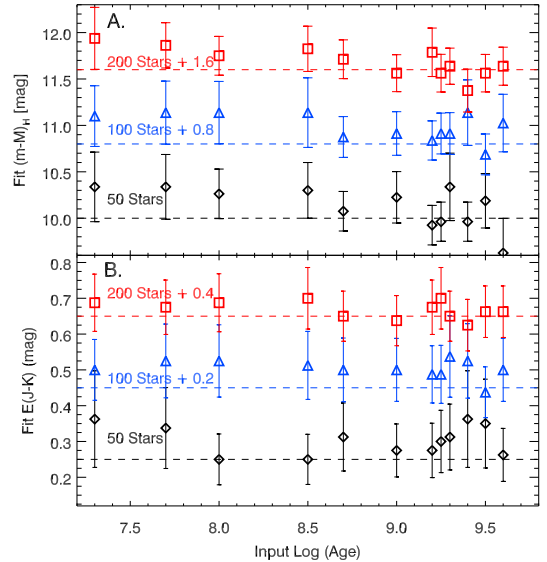}
		\caption{The best-fit distance modulus (A, Top, uncorrected for reddening) and E$(J-K)$ (B, Bottom) of the synthetic clusters are plotted against input log(age) of the synthetic clusters.  The 50-star, 100-star, and 200-star clusters are shown as black diamonds, blue triangles, and red squares, respectively.  The 100-star and 200-star clusters were displaced vertically by 0.8 and 1.6 mag, respectively, for distance modulus, and by 0.2 and 0.4 mag, respectively for color excess.  The dashed lines show the input parameters of 10 mag for (m-M)$_H$ and 0.25 mag for E$(J-K)$.}
		\label{synclusterdistcolfig}
	\end{center}
\end{figure}

Figure \ref{synclusterdistcolfig} (A, top) shows the distance modulus returned from the isochrone fitting as a function of input age.  The input distance modulus (10 mag in H-band) is marked by the dashed lines (with offsets).  Similarly, for E$(J-K)$, shown in Figure \ref{synclusterdistcolfig} (B, bottom), the input E$(J-K)$ of 0.25 mag is indicated by dashed lines.  For both distance modulus and color excess, larger deviations are seen mostly at the youngest ages.  We suspect these deviations are due to the absence of a red clump in the cluster CMDs, which leaves the fit somewhat unconstrained.  More than 85\% of the fit distance modulus values, and all the fit color excess values, fall within 1$\sigma$ of their input values.  


Based on these analyses of fitting to synthetic clusters, for which we know the true parameters a priori, we conclude that the overall isochrone fitting procedure is reliable within the uncertainties it returns, provided that the cluster membership is not overly sparse.  Additionally, the prevalence of deviations that are less than 1$\sigma$ may indicate that the calculated uncertainties are somewhat overestimated.  We elect to retain them, though, as conservative uncertainty estimates.  

\section{Discussion}\label{Discussion}
To better understand the properties of the cluster sample, we searched for trends among the cluster parameters.  Next, we compared the results of the present study to those found in previous studies to determine whether, and to what degree, our findings differ from previous ones.

\subsection{The Properties of the 30 Clusters fit to Isochrones}  
The properties of the 30 clusters, determined via fitting theoretical isochrones, span wide ranges of age, distance, and reddening.  The standard deviations of the distributions of the best-fit log(age), distances, and E$(B-V)$ are 0.63, 1400 pc, and 0.3 mag, respectively, and the median values are  $\sim$9.2, 2900 pc, and 0.5 mag, respectively.  The farthest cluster is Berkeley 60, at just over 6 kpc, and the nearest cluster is M 67 at $\sim$670 pc.

The subsample of six clusters for which only distance modulus and color excess were fit have relatively larger distance and reddening uncertainties than the subsample of 24 clusters for which all three parameters were fit.  The mean uncertainties for the six clusters were $\sim$420 pc and 0.15 mag for distance and E(B-V), respectively.  For the 24 clusters, the mean uncertainties were $\sim$320 pc, 0.14 mag, and 0.13 for distance, E(B-V), and log(age), respectively.

The six clusters have somewhat larger parameter uncertainties due to their sparse and/or faint nature.  As can be seen in Figure \ref{rejcmdbestfitfig}, Basel 11b and NGC 2126 have few cluster members in their CMDs, $\sim$30 stars, which may not be enough to reliably constrain the fit.  The four clusters Berkeley 14, Berkeley 18, Berkeley 39, and Berkeley 32 all have main sequence turnoffs near $\sim$14--14.5 mag in $H$-band, just above our $H$ = 15 mag limit.  This does not provide an adequate portion of the cluster main sequence to be fit.

Based on the numbers of CMD stars of the clusters that could not be fully fit, and from the isochrone fits to the synthetic clusters, we expect that the fitting procedure is less robust for clusters with fewer than $\sim$50 stars in their CMDs or whose main sequence turnoff is below $H\sim$14 mag.  For these faint clusters, deeper photometric data would enable more successful fits.

We note that NGC 2266, which was part of the 24 successfully fit subsample, has fewer than 50 stars in its CMD.  Some caution may be advised when adopting its parameters from this study.

\subsubsection{Comparison of Derived Cluster Properties}
\begin{figure}  
	\begin{center}
		\includegraphics[width= 0.45\textwidth]{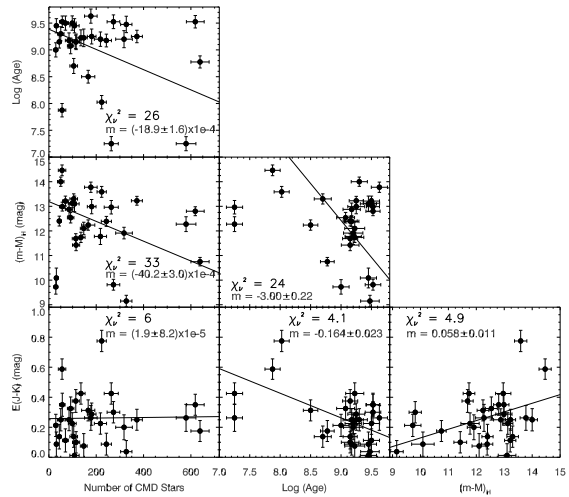}
		\caption{Best-fitting parameters of the 30 clusters whose CMDs were reliably fit, plotted against the other parameters to search for trends.  (Left column.) Log (age), (m-M)$_H$ , and E$(J-K)$ plotted against number of CMD stars.  (Middle column.) (m-M)$_H$ and E$(J-K)$ plotted against log(age).  (Right column.) E$(J-K)$ plotted against (m-M)$_H$.  Linear fits to each set of values are overlaid as black solid lines, and the slope (m) and reduced $\chi^2$ value of each fit are reported.}
		\label{outputagedistredfig}
	\end{center}
\end{figure}

The derived parameters of the 30 clusters were searched for correlations for potential biases in the fitting procedure.  Figure \ref{outputagedistredfig} plots the best-fit log(age), (m-M)$_H$, and E$(J-K)$ values against each other and against the number of CMD members.  A linear fit was computed for each comparison, and the slope of each fit (m), along with its reduced $\chi^2$ value, are listed in each plot.
 
No obvious trends exist between the number of CMD stars and the three fit parameters.  There is a correlation between the distance modulus and color excess, which is reasonable given that a cluster is likely to be more extincted if it is farther away.  Both distance modulus and color excess appear to decrease as a function of increasing cluster age.  The trend of distance modulus decreasing with age may be because younger clusters contain relatively brighter stars that can be seen at larger distances.  The correlation between age and color excess may be due to the likelihood of finding older clusters farther from the Galactic plane \citep{Friel1995}, which would be along less extincted sight-lines.

\subsection{Comparison to Previous Studies}
We compared our derived cluster parameters to those found in recent publications to determine whether there were significant differences.  Table \ref{clusterlitcomparison} summarizes the relevant parameters found in the previous studies for the 30 clusters.  The fifth column of the Table lists whether our cluster log(ages), distances (in pc), and reddenings (E$(B-V)$) agree with those in the cited sources to within 1$\sigma$, 2$\sigma$, 3$\sigma$, or $>$ 3$\sigma$ (a, b, c, d labels, respectively), based on our uncertainty estimates for each parameter.  Log(age), distance in pc, and E$(B-V)$ values were used for comparisons because these were the most cited properties in the literature, especially as most of the studies were done in the optical.  We find that for most of the parameters of the majority of clusters ($\sim$80\% of all the cluster parameters), the derived values agree to within 2$\sigma$.  As no consensus of concordance was found in the literature for comparisons of the agreement of derived cluster properties, we adopted 2$\sigma$ as the standard of agreement.  

\begin{figure}  
	\begin{center}
		\includegraphics[width= 0.45\textwidth]{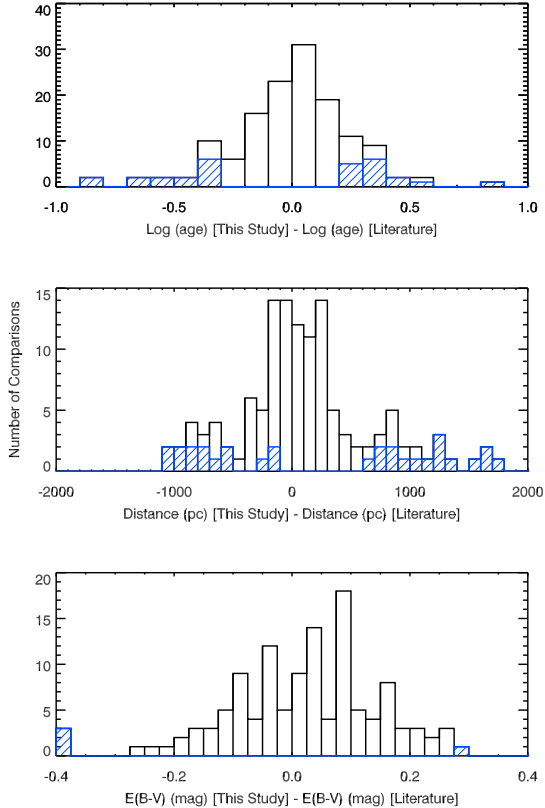}
		\caption{The distributions of differences between the cluster parameters derived in the present study and those derived in previous studies.  The blue striped distributions indicate the numbers of comparisons that differed by greater than 2$\sigma$.  One point in the age distribution, four points in the distance distribution, and two in the E$(B-V)$ distribution are not shown, as they are outside the plot ranges.}
		\label{litdifferencefig}
	\end{center}
\end{figure}

Figure \ref{litdifferencefig} plots the distributions of the differences between the parameters derived in this study to those of previous studies.  No systematic offsets were found between our values of log(age), distance, and reddening and those cited.  Reddening estimates show the most agreement ($>$ 95\%) with previous values, while the derived ages and distances agree to within our uncertainties for $\ge$85\% of the cited values.

\subsection{Galactic Locations and Cluster Properties}
\begin{figure} 
	\begin{center}
		\includegraphics[width= 0.45\textwidth]{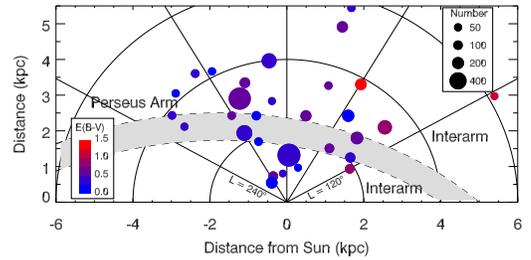}
		\caption{Locations of the 30 fit clusters in the outer Galaxy, with the Sun at the origin.  The clusters are shown as filled circles, and the color denotes the degree of reddening, as indexed by the legend in the bottom left corner.  The sizes of the circles correspond to the number of cluster CMD members (Table \ref{clusterbestfittable}), as indicated by the legend in the upper right corner.  The approximate location of the Perseus Arm \citep{Reid2014} is shown as the grey stripe.}
		\label{mapclustersfig}
	\end{center}
\end{figure}

Figure \ref{mapclustersfig} plots the locations of the 30 fit clusters, with the Sun at the origin, using the derived distances listed in Table \ref{clusterbestfittable}.  The Perseus Spiral Arm \citep{Reid2014} is denoted as the grey stripe.  The clusters span a wide range of distances, residing in both arm and interarm locations.  The clusters located beyond the Perseus Arm have slightly larger reddenings on average than the clusters located in front of the arm.  These distances will be combined with NIR polarimetry of the cluster stars to probe the nature of the magnetic field in the outer Galaxy.

\section{Summary}
To study the large-scale structure of the Galactic magnetic field in the outer Galaxy, we obtained NIR stellar polarimetric measurements of a sample of 31 open clusters spanning wide ranges of longitude, distance, and reddening.  It was essential to determine the cluster distances accurately to optimally use the polarimetric information they provided.  To do so, we developed a $\chi^2$ minimization technique to fit theoretical isochrones to the cluster CMDs.  These fits returned the cluster properties of distance, age, and reddening, along with their corresponding uncertainties.

For each cluster, the field-star contamination was removed from the cluster CMD, and Monte Carlo realizations of this background-subtracted cluster CMD were created.  Monte Carlo isochrone realizations, based on the PARSEC isochrones, were created at different ages, distance moduli and color excesses to compare to each cluster CMD.  The isochrone parameters which yielded the lowest color-magnitude distance-based mean $\chi^2$ were adopted as the cluster parameters.  Of the original sample of 31 clusters, 24 were fit for all three parameters.  The remaining seven were either faint or sparse.  By fixing age, distance and reddening estimates were fit for six of the seven.  The mean uncertainties of the 30 clusters of log(age), distance, and $E(B-V)$ were 0.13, 340 pc, and 0.14 mag, respectively.

To test the accuracy of the fitting technique, synthetic clusters were created and fit.  These clusters were generated at 12 ages, ranging in log(age) from 7.3 to 9.6, for 50, 100, and 200 numbers of stellar members.  For nearly all of the synthetic clusters, the input parameters were recovered to within their 1$\sigma$ uncertainties.  The clusters with 50 members were the most difficult to fit and showed the largest scatter from their input values, though a strong majority were still fit to within 1$\sigma$ of their input parameters.

The derived properties of the 30 clusters that were fit were compared to values found in recent published studies and revealed no biases or trends.  

This sample of clusters spans wide ranges in longitude and distance.  It also probes the Perseus Spiral arm, as well as its foreground and background interarm regions.  The cluster distribution is well suited to probe the properties of the large-scale Galactic magnetic field in the outer Galaxy and to test whether the field is affected by the presence of a spiral arm.

\section{Acknowledgments}
This research made use of the NASA/ IPAC Infrared Science Archive, which is operated by the Jet Propulsion Laboratory, California Institute of Technology (Caltech), under contract with NASA.  This publication made use of data products from the Two Micron All Sky Survey, which was a joint project of the University of Massachusetts and the Infrared Processing and Analysis Center/Caltech, funded by NASA and the National Science Foundation (NSF).  We thank the anonymous referee for providing constructive comments and suggestions.  We also thank April Pinnick for selecting the cluster sample and conducting NIR observations of many of these clusters.  We are grateful to A. A. West, L. Cashman, and J. Montgomery for providing valuable feedback on the manuscript. This work was supported under NSF grants AST 09-07790 and 14-12269 and NASA grant NNX15AE51G to Boston University.




\newpage
\begin{deluxetable}{cccccccc}   
\tabletypesize{\footnotesize}
\tablecolumns{8}
\tablecaption{Open Cluster Sample and Stellar Radial Density Profile Parameters}
\tablewidth{0pt}
\tablehead{
 \colhead{Name} &
 \colhead{$l$} &
 \colhead{$b$} &
 \colhead{Number of} &
 \colhead{R$_{1.5GW}$ \tablenotemark{a}} &
 \colhead{Background} &
 \colhead{Metallicity} &
 \colhead{Metallicity} \\
 \colhead{} &
 \colhead{(deg)} &
 \colhead{(deg)} &
 \colhead{Stars} &
 \colhead{(arcsec)} &
 \colhead{(Stars arcmin$^{-2}$)} &
 \colhead{(Z)} &
 \colhead{Ref.\tablenotemark{b}} 
}
\startdata
Berkeley 60 &   118.846 &    -1.643 & 130 $\pm$50 & 180 $\pm$29 & 6.21 $\pm$0.07 & 0.019 & --- \\
King 1 &   119.763 &     1.693 & 220 $\pm$50 & 161 $\pm$15 & 6.96 $\pm$0.07 & 0.019 & --- \\
NGC 559 &   127.194 &     0.746 & 360 $\pm$80 & 275 $\pm$25 & 6.32 $\pm$0.07 & 0.019 & --- \\
NGC 663 &   129.467 &    -0.941 & 520 $\pm$100 & 318 $\pm$25 & 6.72 $\pm$0.08 & 0.019 & --- \\
NGC 869 &   134.633 &    -3.752 & 1020 $\pm$150 & 455 $\pm$27 & 5.47 $\pm$0.09 & 0.019 & --- \\
King 5 &   143.773 &    -4.276 & 220 $\pm$50 & 177 $\pm$15 & 4.07 $\pm$0.06 & 0.008 & \citep{Friel1995} \\
NGC 1245 &   146.654 &    -8.908 & 430 $\pm$70 & 280 $\pm$18 & 3.27 $\pm$0.05 & 0.017 & \citep{Jacobson2011} \\
King 7 &   149.799 &    -1.021 & 490 $\pm$70 & 201 $\pm$11 & 5.02 $\pm$0.06 & 0.019 & --- \\
Berkeley 12 &   161.677 &    -1.992 & 170 $\pm$40 & 145 $\pm$14 & 4.64 $\pm$0.06 & 0.019 & --- \\
Berkeley 14 &   162.873 &     0.707 & 170 $\pm$50 & 174 $\pm$19 & 5.21 $\pm$0.06 & 0.019 & --- \\
NGC 2126 &   163.236 &    13.140 & 54 $\pm$24 & 124 $\pm$21 & 2.11 $\pm$0.04 & 0.019 & --- \\
Berkeley 18 &   163.648 &     5.050 & 410 $\pm$100 & 410 $\pm$40 & 3.91 $\pm$0.07 & 0.019 & --- \\
Berkeley 70 &   166.902 &     3.585 & 200 $\pm$70 & 300 $\pm$40 & 4.11 $\pm$0.06 & 0.019 & \citep{Carrera2012} \\
NGC 1857\tablenotemark{c} &   168.444 &     1.222 & 360 $\pm$100 & 390 $\pm$40 & 4.97 $\pm$0.07 & 0.005 & --- \\
NGC 2099 &   177.636 &     3.092 & 1030 $\pm$140 & 448 $\pm$24 & 5.12 $\pm$0.08 & 0.019 & \citep{Pancino2010} \\
NGC 2158 &   186.634 &     1.781 & 890 $\pm$80 & 192 $\pm$6 & 5.87 $\pm$0.07 & 0.010 & \citep{Jacobson2011} \\
Basel 11b &   187.442 &    -1.114 & 32 $\pm$21 & 90 $\pm$24 & 4.88 $\pm$0.06 & 0.019 & --- \\
NGC 2266 &   187.778 &    10.304 & 88 $\pm$25 & 94 $\pm$9 & 2.53 $\pm$0.04 & 0.007 & \citep{Reddy2013} \\
NGC 2141 &   198.044 &    -5.810 & 440 $\pm$60 & 218 $\pm$12 & 3.59 $\pm$0.05 & 0.019 & \citep{Jacobson2009} \\
NGC 2420 &   198.107 &    19.634 & 360 $\pm$50 & 225 $\pm$12 & 1.45 $\pm$0.04 & 0.012 & \citep{Jacobson2011} \\
Trumpler 5 &   202.807 &     1.018 & 1510 $\pm$140 & 434 $\pm$16 & 4.59 $\pm$0.08 & 0.008 & \citep{Carrera2007} \\
NGC 2355 &   203.390 &    11.803 & 230 $\pm$50 & 246 $\pm$21 & 2.21 $\pm$0.04 & 0.016 & \citep{Jacobson2011} \\
NGC 2112 &   205.872 &   -12.615 & 420 $\pm$100 & 470 $\pm$50 & 2.20 $\pm$0.06 & 0.015 & \citep{Brown1996} \\
Berkeley 32 &   207.952 &     4.404 & 280 $\pm$70 & 283 $\pm$28 & 3.82 $\pm$0.06 & 0.0095 & \citep{Carrera2011} \\
Collinder 110 &   209.649 &    -1.927 & 530 $\pm$120 & 460 $\pm$40 & 4.66 $\pm$0.08 & 0.019 & \citep{Carrera2007} \\
NGC 2262 &   210.573 &    -2.099 & 190 $\pm$50 & 151 $\pm$14 & 4.42 $\pm$0.06 & 0.019 & --- \\
NGC 2324 &   213.447 &     3.297 & 210 $\pm$50 & 198 $\pm$19 & 4.39 $\pm$0.06 & 0.0128 & \citep{Bragaglia2008} \\
M 67 &   215.696 &    31.896 & 450 $\pm$90 & 520 $\pm$40 & 1.03 $\pm$0.04 & 0.019 & \citep{Jacobson2011} \\
Berkeley 39 &   223.461 &    10.095 & 210 $\pm$40 & 186 $\pm$15 & 2.34 $\pm$0.04 & 0.014 & \citep{Carrera2007} \\
Haffner 10 &   230.799 &     1.011 & 250 $\pm$60 & 185 $\pm$16 & 5.49 $\pm$0.07 & 0.019 & --- \\
NGC 2425 &   231.501 &     3.289 & 150 $\pm$50 & 184 $\pm$23 & 4.28 $\pm$0.06 & 0.0135 & \citep{Jacobson2011} \\
\enddata
\label{clustercoordinatestable}
\tablenotetext{a}{R$_{1.5GW}$ signifies the radius of the inner region, which is equal to 1.5 times the Gaussian width of the cluster radial density profile.}
\tablenotetext{b}{For clusters where no spectroscopic metallicity measurement was found, no reference is listed.}
\tablenotetext{c}{For cluster NGC 1857, a metallicity of Z = 0.005 was assumed instead of solar metallicity.}
\end{deluxetable}

\clearpage

\begin{deluxetable}{ccccccc}  
\tabletypesize{\footnotesize}
\tablecolumns{7}
\tablecaption{Derived Cluster Properties}
\tablewidth{0pt}
\tablehead{
\colhead{Name} &
\colhead{Number of \tablenotemark{a}} &
\colhead{Log (age)} &
\colhead{(m-M)$_H$} &
\colhead{E$(J-K)$} &
\colhead{Distance} &
\colhead{E$(B-V$)\tablenotemark{c}}\\
\colhead{ } &
\colhead{CMD Stars} &
\colhead{[log(yrs)]} &
\colhead{(mag)} &
\colhead{(mag)} &
\colhead{(pc)} &
\colhead{(mag)}
}
\startdata	
Berkeley 60 & 56 $\pm$16 & 7.88 $\pm$0.13 & 14.46 $\pm$0.22 & 0.59 $\pm$0.07 & 6160 $^{+680}_{-610}$ & 1.11 \\
King 1 & 135 $\pm$17 & 9.22 $\pm$0.15 & 11.74 $\pm$0.21 & 0.42 $\pm$0.07 & 1880 $^{+200}_{-180}$ & 0.80 \\
NGC 559 & 218 $\pm$26 & 9.20 $\pm$0.13 & 11.80 $\pm$0.30 & 0.23 $\pm$0.07 & 2070 $^{+340}_{-290}$ & 0.42 \\
NGC 663 & 260 $\pm$30 & 7.25 $\pm$0.13 & 12.96 $\pm$0.27 & 0.42 $\pm$0.07 & 3300 $^{+460}_{-400}$ & 0.80 \\
NGC 869 & 580 $\pm$40 & 7.25 $\pm$0.13 & 12.30 $\pm$0.30 & 0.26 $\pm$0.09 & 2560 $^{+400}_{-340}$ & 0.50 \\
King 5 & 114 $\pm$15 & 9.15 $\pm$0.12 & 11.69 $\pm$0.20 & 0.38 $\pm$0.08 & 1870 $^{+190}_{-170}$ & 0.71 \\
NGC 1245 & 243 $\pm$21 & 9.18 $\pm$0.12 & 12.39 $\pm$0.20 & 0.09 $\pm$0.07 & 2900 $^{+300}_{-270}$ & 0.17 \\
King 7 & 223 $\pm$20 & 8.03 $\pm$0.12 & 13.59 $\pm$0.22 & 0.78 $\pm$0.07 & 3820 $^{+430}_{-380}$ & 1.46 \\
Berkeley 12 & 57 $\pm$12 & 9.53 $\pm$0.11 & 12.99 $\pm$0.18 & 0.35 $\pm$0.08 & 3440 $^{+320}_{-290}$ & 0.66 \\
Berkeley 14 & 50 $\pm$14 & 9.30\tablenotemark{b} & 14.00 $\pm$0.19 & 0.25 $\pm$0.07 & 5700 $^{+550}_{-500}$ & 0.47 \\
NGC 2126 & 32 $\pm$8 & 9.45\tablenotemark{b} & 10.10 $\pm$0.40 & 0.09 $\pm$0.08 & 1010 $^{+210}_{-170}$ & 0.17 \\
Berkeley 18 & 177 $\pm$27 & 9.63\tablenotemark{b} & 13.78 $\pm$0.21 & 0.26 $\pm$0.08 & 5120 $^{+550}_{-500}$ & 0.50 \\
NGC 1857 & 166 $\pm$28 & 8.50 $\pm$0.12 & 12.24 $\pm$0.21 & 0.31 $\pm$0.07 & 2470 $^{+260}_{-240}$ & 0.59 \\
NGC 2099 & 640 $\pm$40 & 8.78 $\pm$0.11 & 10.75 $\pm$0.17 & 0.17 $\pm$0.07 & 1320 $^{+120}_{-110}$ & 0.33 \\
NGC 2158 & 372 $\pm$23 & 9.25 $\pm$0.12 & 13.23 $\pm$0.18 & 0.25 $\pm$0.07 & 3990 $^{+370}_{-330}$ & 0.47 \\
Basel 11b & 29 $\pm$8 & 9.00\tablenotemark{b} & 9.72 $\pm$0.29 & 0.21 $\pm$0.07 & 810 $^{+120}_{-100}$ & 0.40 \\
NGC 2266 & 44 $\pm$8 & 9.15 $\pm$0.12 & 12.40 $\pm$0.20 & 0.14 $\pm$0.08 & 2860 $^{+290}_{-270}$ & 0.26 \\
NGC 2141 & 181 $\pm$18 & 9.25 $\pm$0.14 & 12.99 $\pm$0.29 & 0.29 $\pm$0.07 & 3520 $^{+520}_{-450}$ & 0.54 \\
NGC 2420 & 147 $\pm$14 & 9.22 $\pm$0.16 & 12.10 $\pm$0.25 & 0.07 $\pm$0.08 & 2550 $^{+330}_{-290}$ & 0.14 \\
Trumpler 5 & 620 $\pm$40 & 9.53 $\pm$0.12 & 12.80 $\pm$0.18 & 0.35 $\pm$0.07 & 3150 $^{+290}_{-270}$ & 0.66 \\
NGC 2355 & 115 $\pm$15 & 9.15 $\pm$0.14 & 11.43 $\pm$0.23 & 0.10 $\pm$0.07 & 1850 $^{+210}_{-190}$ & 0.19 \\
NGC 2112 & 273 $\pm$27 & 9.53 $\pm$0.13 & 9.81 $\pm$0.22 & 0.30 $\pm$0.07 & 810 $^{+90}_{-80}$ & 0.57 \\
Berkeley 32 & 108 $\pm$20 & 9.45\tablenotemark{b} & 13.10 $\pm$0.30 & 0.01 $\pm$0.08 & 4150 $^{+670}_{-570}$ & 0.02 \\
Collinder 110 & 320 $\pm$40 & 9.20 $\pm$0.16 & 11.91 $\pm$0.24 & 0.20 $\pm$0.08 & 2230 $^{+270}_{-240}$ & 0.38 \\
NGC 2262 & 92 $\pm$14 & 9.07 $\pm$0.15 & 12.54 $\pm$0.21 & 0.33 $\pm$0.08 & 2820 $^{+310}_{-280}$ & 0.61 \\
NGC 2324 & 105 $\pm$16 & 8.70 $\pm$0.14 & 13.30 $\pm$0.21 & 0.14 $\pm$0.07 & 4320 $^{+450}_{-410}$ & 0.26 \\
M 67 & 327 $\pm$24 & 9.47 $\pm$0.15 & 9.15 $\pm$0.23 & 0.04 $\pm$0.07 & 670 $^{+80}_{-70}$ & 0.07 \\
Berkeley 39 & 70 $\pm$12 & 9.50\tablenotemark{b} & 13.21 $\pm$0.20 & 0.11 $\pm$0.08 & 4200 $^{+430}_{-390}$ & 0.21 \\
Haffner 10 & 99 $\pm$17 & 9.50 $\pm$0.13 & 13.13 $\pm$0.18 & 0.23 $\pm$0.07 & 3850 $^{+360}_{-330}$ & 0.42 \\
NGC 2425 & 85 $\pm$16 & 9.18 $\pm$0.14 & 12.88 $\pm$0.26 & 0.25 $\pm$0.07 & 3400 $^{+440}_{-390}$ & 0.47 \\
\enddata
\label{clusterbestfittable}
\tablenotetext{a}{The numbers of CMD stars are the total number of star counts of the cluster CMD overdensity images, and represent the counts the Poisson images are drawn from.  See Section \ref{results of isochrone fitting} for discussion.}
\tablenotetext{b}{Uncertainties in log(age) are not listed, as these clusters were only fit for distance modulus and color excess, and their ages were fixed at the values listed in \citet{Kharchenko2013}.}
\tablenotetext{c}{The uncertainties of E$(B-V)$ are equal to the uncertainties of E$(J-K)$/0.53, following the relation between E$(B-V)$ and E$(J-K)$.}
\end{deluxetable}

\newpage



\clearpage

\LongTables
\begin{deluxetable*}{cccccc} 
\tabletypesize{\scriptsize}
\tablecolumns{6}
\tablecaption{Cluster Parameters Found in Recent Studies}
\tablewidth{0pt}
\tablehead{
\colhead{Name} &
\colhead{Log (age)} &
\colhead{D} &
\colhead{E(B-V)} &
\colhead{Agreement\tablenotemark{a}} &
\colhead{Ref.} \\
\colhead{} &
\colhead{[log(yrs)]} &
\colhead{(pc)} &
\colhead{(mag)} & 
\colhead{} &
\colhead{}
}
\startdata
Berkeley 60 & 8.25 & 4468 & 0.92 & ccb & \citet{Bukowiecki2011} \\
Berkeley 60 & 8.2 & 4365 & 0.86 & ccb & \citet{Ann2002b} \\
Berkeley 60 & 8.4 & 3299 & 0.958 & ddb & \citet{Kharchenko2013} \\
Berkeley 60 & 8.2 & 2089 & 0.37 & cdd & \citet{Tadross2001} \\
King 1 & 9.45 & 2060 & 0.625 & bab & \citet{Hasegawa2008} \\
King 1 & 9.6 & 1080 & 0.76 & cda & \citet{Maciejewski2007} \\
King 1 & 9.2 & 1900 & 0.7 & aaa & \citet{Lata2004} \\
King 1 & 9.7 & 1659 & 0.625 & dbb & \citet{Kharchenko2013} \\
NGC 559 & 8.35 & 2430 & 0.82 & dbd & \citet{Joshi2014} \\
NGC 559 & 8.8 & 2170 & 0.68 & dab & \citet{Maciejewski2007} \\
NGC 559 & 8.6 & 2291 & 0.81 & dac & \citet{Ann2002a} \\
NGC 559 & 7.7 & 6309 & 0.62 & ddb & \citet{Jennens1975} \\
NGC 559 & --- & 1200 & --- & -c- & \citet{Grubissich1975} \\
NGC 559 & 9.08 & 1300 & --- & ac- & \citet{Lindoff1969} \\
NGC 559 & 8.8 & 2200 & 0.6 & dab & \citet{Kharchenko2013} \\
NGC 663 & 7.65 & 2520 & 0.63 & dbb & \citet{Bukowiecki2011} \\
NGC 663 & 7.3-7.4\tablenotemark{b} & 2420 & 0.8 & aba & \citet{Pandey2005} \\
NGC 663 & 7.4 & 2089 & --- & bc- & \citet{Fabregat2005} \\
NGC 663 & 7.3 & 2469 & 0.75 & aba & \citet{Tadross2001} \\
NGC 663 & 7.3-7.4 & 2100 & 0.83 & aca & \citet{Pigulski2001} \\
NGC 663 & 7.86 & 1718 & --- & dd- & \citet{Malysheva1997} \\
NGC 663 & 7.08-7.4 & 2818 & 0.8 & aba & \citet{Phelps1994} \\
NGC 663 & 7.5 & 2100 & 0.7 & bca & \citet{Kharchenko2013} \\
NGC 869 & 7.14 & 2290 & 0.55 & aaa & \citet{Currie2010} \\
NGC 869 & 7.11 & 2269 & 0.54 & baa & \citet{Mayne2008} \\
NGC 869 & --- & --- & 0.52 & --a & \citet{Bragg2005} \\
NGC 869 & 7.11 & 2344 & 0.56 & baa & \citet{Slesnick2002} \\
NGC 869 & 7.1 & 2188 & --- & ba- & \citet{Capilla2002} \\
NGC 869 & 7.1 & 2904 & 0.58 & baa & \citet{Tadross2001} \\
NGC 869 & 6.91 & 2025 & --- & cb- & \citet{Malysheva1997} \\
NGC 869 & 7.28 & 2300 & 0.521 & aaa & \citet{Kharchenko2013} \\
King 5 & 9.1 & 2230 & 0.67 & aba & \citet{Maciejewski2007} \\
King 5 & 9 & 1900 & 0.82 & baa & \citet{Durgapal2001} \\
King 5 & 9 & 1905 & 0.94 & bab & \citet{Carraro2000} \\
King 5 & 8.9 & --- & --- & c-- & \citet{Salaris2004} \\
King 5 & 9.09 & 2200 & 0.67 & aba & \citet{Kharchenko2013} \\
NGC 1245 & 9.03 & 2818 & 0.24 & baa & \citet{Lee2012} \\
NGC 1245 & 9.04 & 3010 & 0.05 & baa & \citet{Alves2012} \\
NGC 1245 & 9.02 & 2800 & --- & ba- & \citet{Burke2004} \\
NGC 1245 & 8.95 & 3019 & 0.29 & baa & \citet{Subramaniam2003} \\
NGC 1245 & 9.03 & --- & --- & b-- & \citet{Salaris2004} \\
NGC 1245 & 9.16 & 2211 & 0.27 & aca & \citet{Tadross2001} \\
NGC 1245 & 9.025 & 3000 & 0.25 & baa & \citet{Kharchenko2013} \\
King 7 & 8.7 & --- & --- & d-- & \citet{Durgapal2001} \\
King 7 & 8.78-8.9 & 2200 & 1.25 & ddb & \citet{Durgapal1997} \\
King 7 & 8.6 & 2440 & 1.25 & ddb & \citet{Tadross2001} \\
King 7 & 8.85 & 2628 & 1.249 & dcb & \citet{Kharchenko2013} \\
Berkeley 12 & 9.2 & 3801 & 0.8 & dba & \citet{Hasegawa2004} \\
Berkeley 12 & 9.6 & 3162 & 0.7 & aaa & \citet{Ann2002b} \\
Berkeley 12 & 9.6 & 3300 & 0.7 & aaa & \citet{Kharchenko2013} \\
NGC 1857 & 9 & 1400 & 0.13 & ddd & \citet{Zasowski2013} \\
NGC 1857 & 8.0-8.25 & 5750 & 0.38-0.6 & dda & \citet{Sujatha2006} \\
NGC 1857 & 8.2 & 1545 & 0.97 & cdc & \citet{Tadross2011} \\
NGC 1857 & 8.67 & 3299 & 0.5 & bda & \citet{Kharchenko2013} \\
NGC 2099 & 8.34-8.51 & --- & --- & d-- & \citet{Salaris2009} \\
NGC 2099 & 8.69 & 1490 & 0.227 & aba & \citet{Hartman2008} \\
NGC 2099 & --- & 1905 & 0.21 & -da & \citet{Kang2007} \\
NGC 2099 & 8.8 & 1995 & 0.23 & ada & \citet{Kalirai2005} \\
NGC 2099 & 8.6-8.72 & 1148-1202 & 0.36 & bba & \citet{Kalirai2004} \\
NGC 2099 & 8.55 & 1400 & 0.35 & caa & \citet{Kharchenko2013} \\
NGC 2158 & 9.28 & 3944 & 0.42 & aaa & \citet{Bedin2010} \\
NGC 2158 & 9.28 & --- & --- & a-- & \citet{Salaris2004} \\
NGC 2158 & 9.3 & 3600 & 0.55 & aba & \citet{Carraro2002} \\
NGC 2158 & --- & 4068 & 0.43 & -aa & \citet{Grocholski2002} \\
NGC 2158 & 9.2 & 5012 & 0.4 & aca & \citet{Tadross2001} \\
NGC 2158 & 9.33 & 4770 & 0.333 & acb & \citet{Kharchenko2013} \\
NGC 2266 & 9 & 2855 & 0.21 & baa & \citet{Dias2012} \\
NGC 2266 & 9.08 & 2800 & 0.17 & aaa & \citet{Maciejewski2008} \\
NGC 2266 & 8.94 & --- & --- & b-- & \citet{Salaris2004} \\
NGC 2266 & 8.9 & 3758 & 0.1 & cdb & \citet{Tadross2001} \\
NGC 2266 & 9.265 & 3311 & 0 & abb & \citet{Kharchenko2013} \\
NGC 2141 & 9.1-9.28 & 4090-4370 & 0.36-0.45 & abb & \citet{Donati2014a} \\
NGC 2141 & 9.4 & --- & --- & b-- & \citet{Salaris2004} \\
NGC 2141 & 9.4 & 3800 & 0.4 & bab & \citet{Carraro2001} \\
NGC 2141 & 9.4 & 4200 & 0.35 & bbb & \citet{Rosvick1995} \\
NGC 2141 & 9.245 & 4364 & 0.312 & abb & \citet{Kharchenko2013} \\
NGC 2420 & 9.3 & 2480 & 0.04 & aaa & \citet{Sharma2006} \\
NGC 2420 & 9.3 & 2542 & 0.04 & aaa & \citet{Mermilliod2007} \\
NGC 2420 & 9.34 & 2443 & 0.05 & aaa & \citet{Salaris2004} \\
NGC 2420 & 9.3 & 2449 & 0.05 & aaa & \citet{Grocholski2003} \\
NGC 2420 & 9.365 & 2880 & 0.01 & aaa & \citet{Kharchenko2013} \\
Trumpler 5 & 9.7 & 2400 & 0.5 & bcb & \citet{Piatti2004b} \\
Trumpler 5 & 9.54-9.6 & 2818-3076 & 0.6-0.7 & aaa & \citet{Donati2014b} \\
Trumpler 5 & 9.45 & 3100 & 0.64 & aaa & \citet{Kim2009} \\
Trumpler 5 & 9.75 & --- & --- & b-- & \citet{Salaris2004} \\
Trumpler 5 & 9.61 & 3019 & 0.58 & aaa & \citet{Kaluzny1998} \\
Trumpler 5 & 9.1 & 2958 & 0.58 & daa & \citet{Tadross2001} \\
Trumpler 5 & 9.5 & 2753 & 0.625 & aba & \citet{Kharchenko2013} \\
Trumpler 5 & 9.5 & 2900 & 0.6 & aaa & \citet{Perren2015} \\
NGC 2355 & 8.9 & 1985 & 0.3 & baa & \citet{Dias2012} \\
NGC 2355 & 8.9 & --- & --- & b-- & \citet{Salaris2004} \\
NGC 2355 & 9 & 1650 & 0.16 & baa & \citet{Soubiran2000} \\
NGC 2355 & 8.98 & 1915 & 0.112 & baa & \citet{Tadross2001} \\
NGC 2355 & 8.9 & 2128 & 0.187 & bba & \citet{Kharchenko2013} \\
NGC 2112 & 9.23 & 940 & 0.6 & cba & \citet{Carraro2008} \\
NGC 2112 & 9.45 & 813 & 0.6 & aaa & \citet{Tadross2001} \\
NGC 2112 & 9.6 & 750 & 0.6 & aaa & \citet{Richtler1989} \\
NGC 2112 & 9.315 & 977 & 0.625 & bba & \citet{Kharchenko2013} \\
Collinder 110 & 9.08-9.23 & 1949-2187 & 0.52-0.58 & aab & \citet{Bragaglia2003} \\
Collinder 110 & 9.15 & 1950 & 0.5 & aba & \citet{Dawson1998} \\
Collinder 110 & 9.22 & 2362 & 0.416 & aaa & \citet{Kharchenko2013} \\
NGC 2262 & 9 & 3600 & 0.55 & aca & \citet{Carraro2005} \\
NGC 2262 & 8.995 & 2511 & 0.625 & aaa & \citet{Kharchenko2013} \\
NGC 2324 & 8.65 & 3800 & 0.25 & aba & \citet{Piatti2004a} \\
NGC 2324 & 8.83 & --- & --- & a-- & \citet{Salaris2004} \\
NGC 2324 & 8.8 & 4169 & 0.17 & aaa & \citet{Kyeong2001} \\
NGC 2324 & 8.68 & 3842 & 0.239 & aba & \citet{Kharchenko2013} \\
NGC 2324 & 8.8 & 4400 & 0.1 & aab & \citet{Perren2015} \\
M 67 & 9.05 & 722 & 0.24 & cab & \citet{Dias2012} \\
M 67 & 9.6 & 823 & 0.04 & aba & \citet{Balaguer2007} \\
M 67 & 9.56-9.66 & 795 & 0.038 & aba & \citet{VandenBerg2004} \\
M 67 & --- & 766 & 0.038 & -ba & \citet{Laugalys2004} \\
M 67 & 9.6 & 832 & 0.04 & aca & \citet{Sandquist2004} \\
M 67 & 9.51 & 870 & 0 & aca & \citet{Bonatto2003} \\
M 67 & 9.535 & 890 & 0.05 & aca & \citet{Kharchenko2013} \\
Haffner 10 & 9.2-9.4 & 3700 & 0.55 & baa & \citet{Vazquez2010} \\
Haffner 10 & 9.4 & 3100-4300 & 0.41-0.64 & aaa & \citet{Pietrukowicz2006} \\
Haffner 10 & 9.305 & 4873 & 0.5 & bca & \citet{Kharchenko2013} \\
NGC 2425 & 9.56 & 3357 & 0.175 & cac & \citet{Hasegawa2008} \\
NGC 2425 & 9.34 & 3550 & 0.21 & bab & \citet{Moitinho2006} \\
NGC 2425 & 9.4 & 2900-3800 & 0.29 & bab & \citet{Pietrukowicz2006} \\
NGC 2425 & 9.34 & 4330 & 0.21 & bcb & \citet{Kharchenko2013} \\
\enddata
\label{clusterlitcomparison}
\tablenotetext{a}{Quality of agreement within 1, 2, 3, or $>$ 3$\sigma$: a, b, c, d, respectively.}
\tablenotetext{b}{For properties reported as a range, the average of the range was used for comparison.}
\end{deluxetable*}
\end{document}